\documentclass{aa}  
\usepackage{subfig,graphicx}
\usepackage{txfonts}
\graphicspath{{./figures/}}
\usepackage{natbib,twoopt} 
\usepackage[breaklinks=true]{hyperref} 
\bibpunct{(}{)}{;}{a}{}{,} 
\makeatletter
 \newcommandtwoopt{\citeads}[3][][]{\href{http://adsabs.harvard.edu/abs/#3}%
   {\def\hyper@linkstart##1##2{}%
    \let\hyper@linkend\@empty\citealp[#1][#2]{#3}}}    
 \newcommandtwoopt{\citepads}[3][][]{\href{http://adsabs.harvard.edu/abs/#3}%
   {\def\hyper@linkstart##1##2{}%
    \let\hyper@linkend\@empty\citep[#1][#2]{#3}}}      
 \newcommandtwoopt{\citetads}[3][][]{\href{http://adsabs.harvard.edu/abs/#3}%
   {\def\hyper@linkstart##1##2{}%
    \let\hyper@linkend\@empty\citet[#1][#2]{#3}}}      
 \newcommandtwoopt{\citeyearads}[3][][]%
   {\href{http://adsabs.harvard.edu/abs/#3}%
   {\def\hyper@linkstart##1##2{}%
    \let\hyper@linkend\@empty\citeyear[#1][#2]{#3}}}   
\makeatother
\begin{document}
 \title{Fast quasi-periodic oscillations in the eclipsing polar VV Puppis from VLT and XMM-Newton observations}
 
   \titlerunning{Optical QPOs in VV Pup}

\author {J.M. Bonnet-Bidaud\inst{1} 
\and M. Mouchet\inst{2} 
\and E. Falize\inst{3}
\and L. Van Box Som\inst{3}
\and C. Busschaert\inst{3}
\and D.A.H. Buckley     \inst{4}  
\and H.  Breytenbach     \inst{4,5} 
\and  T.R. Marsh \inst{6}
\and R.P. Ashley\inst{6}
\and V.S. Dhillon       \inst{7,8}   
   }
       
   \offprints{J.M. Bonnet-Bidaud}

   \institute{
D\'epartement d'Astrophysique-Laboratoire AIM, CEA/Irfu,  F-91191 Gif-sur-Yvette, France.  (\email{bonnetbidaud@cea.fr)}       
\and LUTH-Observatoire de Paris, UMR 8102-CNRS, Universit\'e Paris-Diderot, F-92190 Meudon, France
\and CEA-DAM-DIF, F-91297 Arpajon, France
 \and  South African Astronomical Observatory, PO Box 9, Observatory 7935, Cape Town, South Africa
 \and Department of Astronomy, University of Cape Town, Rondebosch 7700, Cape Town, South Africa
\and  Department of Physics, University of Warwick, Coventry CV4 7AL, UK
 \and  Department of Physics and Astronomy, University of Sheffield, Sheffield S3 7RH, UK
 \and Instituto de Astrofisica de Canarias, E-38205 La Laguna,  Tenerife, Spain
} 
 
   \date{Received: 23 April 2019; accepted: 12 June 2019}

 
  \abstract{
  We present high time resolution optical photometric data of the polar VV Puppis obtained simultaneously in three filters (u', HeII $\lambda$4686, r') with the ULTRACAM camera mounted at the ESO-VLT telescope. An analysis of a long 50 ks XMM-Newton observation of the source, retrieved from the database, is also provided. 
Quasi-periodic oscillations (QPOs) are clearly detected in the optical during the source bright phase intervals when the accreting pole is visible, confirming the association of the QPOs with the basis of the accretion column.
QPOs are detected in the three filters at a mean frequency of $\sim$ 0.7 Hz with a similar amplitude $\sim$ 1\%. Mean orbitally-averaged power spectra during the bright phase show a rather broad excess with a quality factor Q= $\nu$/$\Delta \nu$ = 5-7 but smaller data segments commonly show a much higher coherency with Q up to 30.
The XMM (0.5--10 keV) observation provides the first accurate estimation of the hard X-ray component with a high kT $\sim$ 40 keV temperature and confirms the high EUV-soft/hard ratio in the range of (4--15) for VV Pup. The detailed X-ray orbital light curve displays a short $\Delta \phi \simeq 0.05$ ingress into self-eclipse of the active pole, indicative of a accretion shock height of $\sim$ 75 km. No significant X-ray QPOs are detected with an amplitude upper limit of $\sim$30\% in the range (0.1--5) Hz.
Detailed hydrodynamical numerical simulations of the post-shock accretion region with parameters consistent with VV Pup demonstrate that the expected frequencies from radiative instability are identical for X-rays and optical regime at values $\nu$ $\sim$ (40--70) Hz, more than one order magnitude higher than observed. This confirms previous statements suggesting that present instability models are unable to explain the full QPO characteristics within the parameters commonly known for polars. 
   }

    \keywords{Physical data and processes:accretion--
    stars:cataclysmic variables --
                stars:white dwarf--
               magnetic white dwarfs -- X-rays: binaries --- accretion ---
               instabilities -- shocks 
               }

   \maketitle
%

\section{Introduction}
The star VV Pup was recognised very early on as a periodic variable star with a 100.4 min period, at the time the shortest known period, from analysis of photographic plates taken at the now defunct Union Observatory (Johannesburg, South Africa)
 \citepads{1931BAN.....6...93V}. 
The binary nature of the source was later established from extended photometry and spectroscopy observations, leading to an accurate ephemeris 
\citepads{1965CoKon..57....1W}, 
 still valid up to this date, more than 50 years later. However, the true nature of the source was only later uncovered with the discovery of strong linear and circular polarization 
\citepads{1977IAUC.3054....1T} 
classifying VV Pup as a member of the AM Her stars, now more commonly known as magnetic cataclysmic variables (MCVs) or Polars.

MCVs are close interacting binaries consisting  of a strongly magnetized ($\gtrsim$10 MG) white dwarf (WD) accreting matter from a red dwarf companion via Roche lobe overflow (see reviews in 
\citeads{2015SSRv..191..111F},  
\citeads{2017PASP..129f2001M}). 
The magnetic field is strong enough to capture matter near L1 and force the flow inside  an accretion column down to the WD magnetic poles. This strong magnetic interaction usually leads to tidally locked systems in which the originally fast WD rotation period slows down to equate with the orbital period, leading to a stable geometry. Material falling down the column is accelerated to a supersonic speed and a shock is formed above the surface of the white dwarf, close to the magnetic pole(s).  Below the shock, the plasma cools by emitting bremsstrahlung in the X-rays and cyclotron radiation in the optical/infrared. 
These two cooling processes lead to a stratification in density and temperature. At the shock, the temperature reaches a few tens of keV, depending mainly on the mass of the white dwarf (see  
\citeads{1990SSRv...54..195C} 
 and  
\citeads{1995CAS....28.....W}). 
According to the exact geometry of the system (orbital inclination, pole colatitude,..), mass transfer may feed one or two of the WD poles. 
 
MCVs are rather "clean" systems in the sense that a unique small emitting region, the bottom of the accretion column below the shock and its immediate surroundings, largely dominate the output radiation. The emission nicely separates into different energy ranges, from infrared and optical (post-shock cyclotron emission), to ultraviolet and soft X-rays (heated white dwarf surface) and to hard X-rays (post-shock thermal free-free emission).

Besides the energy distribution, other interesting diagnostics may come from the observed fast variability. 
Indeed, a sample of MCVs are known to show fast (1--3) s quasi-periodic oscillations (QPOs) that were attributed to hydro-radiative instabilities in the post-shock zone, inducing shock height oscillations with a period of the order of the post-shock cooling time scale 
 (\citeads{1981ApJ...245L..23L},   
 \citeyearads{1982ApJ...258..289L};   
  \citeads{1982ApJ...261..543C}). 
Fast QPOs were discovered by Middleditch in the early 1980s (Middleditch 1982) and currently, five MCVs (V834 Cen, AN UMa, EF Eri, BL Hyi and VV Pup) show (1--5) s, low-amplitude (1--5) \% QPOs in their optical emission.
Coupling the shock energy distribution with the properties of the shock oscillations may theoretically provide a full determination of the major accretion parameters (white dwarf mass, column cross section, accretion rate).
Due to well-controlled physical processes, MCV accretion columns also represent a unique astrophysical environment that can be validly tested by both numerical simulations 
\citepads{2018MNRAS.473.3158V}   
and laser astrophysics experiments 
\citepads{2016NatCo...711899C}.  

However, significant difficulties still exist to unambiguously attribute the observed QPOs to hydro-radiative instabilities. First of all, no X-ray QPOs have been detected so far in Polars whereas high amplitude QPOs are predicted by shock models for an accretion column dominated by bremsstrahlung cooling. A systematic search in the XMM-Newton archive data (0.5--10) keV for some 20 of the brightest polars, did not provide significant detection, with upper limits constraining the column physical parameters 
(\citeads{2015A&A...579A..24B},  
hereafter BB15).
Also, 1D numerical hydrodynamic simulations for a large sample of polar physical parameters have shown that the dominant frequencies of the shock oscillations are generally an order of magnitude greater than the typical QPO detected frequencies 
(\citeads{2015A&A...579A..25B} 
\citeads{2018MNRAS.473.3158V}).   
These inconsistencies show that the standard column model is not complete and that there is a lack of a fundamental physical ingredient.
The reason why only a very limited number of MCVs (5 out of more than one hundred) show these fast QPOs is also not elucidated.

Here, we report on new fast optical photometric observations of VV Pup, obtained at the VLT in order to better characterize the fast quasi-periodic oscillations and directly compare their properties with what expected from the standard accretion model as well as from improved MHD numerical simulations. A similar study was also conducted for the polar V834 Cen 
\citepads{2017A&A...600A..53M}.  
 
VV Pup is one MCV that has two identified poles with surface field of 31.5 MG and 54.6 MG and the magnetic field structure is well described by a 40 MG dipole slightly offset from the WD center 
(\citeads{1989ApJ...342L..35W},   
\citeads{1997AN....318..111S},   
\citeads{2007A&A...467..277M}).   
The binary system consists of a secondary star of mass of (0.10$\pm$0.02) M$_{\sun}$ orbiting a (0.73$\pm$0.05) M$_{\sun}$ WD 
\citepads{2006AJ....131.2216H}.  
From the Gaia mission, the VV Pup distance was recently accurately measured to 137 $\pm 1$ pc 
\citepads{ 2018arXiv180409376L}. 
\\
Only sparse and limited information on the source X-ray light curve and spectrum is available from various previous observations 
(\citeads{1984ApJ...279..785P},   
\citeads{1985xra..conf...63O}, 
\citeads{1996MNRAS.278..285R},   
\citeads{2000PASP..112...18I},   
\citeads{2005ApJ...620..416P}).   
We also present here the first detailed VV Pup X-ray light curve and spectrum during a high state, obtained from an extended XMM-Newton observation. \\
\section{Ultracam optical data}
\subsection{Observations and data reduction}
 VV Pup was observed in May 2005 during two nights (11 and 13 May) at the ESO-VLT telescope using the ULTRACAM camera \citepads{2007MNRAS.378..825D}.  
This ultra-fast CCD camera was mounted as visitor instrument on the UT3 Unit of the 8.2 m Very Large Telescope (VLT).
It is a triple-beam device designed to provide high temporal photometry in three different filters simultaneously. 
The data were acquired using the Sloan Digital Sky Survey filters (SDSS) u' ($\lambda_{eff} $= 3557$\AA$) and r' ($\lambda_{eff}$= 6261$\AA$) and
a narrow-band  filter centered close to the HeII line ($\lambda_{eff}$ = 4662 $\AA$, $108 \AA$ width), noted as He filter in the following. This filter was not active during part of the first night.

A time resolution of 0.097s was chosen to allow the search of QPOs up to 5.1 Hz (Nyquist frequency).
Data were obtained in a two-window drift mode with a 4x4 binning.
Weather conditions were good on May 11 (seeing 0.6"), but less optimal on May 13 with a larger seeing (occurences up to 1.5")  and the occasional presence of cirrus. 
The total exposure time is 5 hr (18 ks). All orbital phases were covered at least three times. 
Table 1 gives a log of the observations.

\begin{table*}
\caption[ ]{VV Pup observing runs with ULTRACAM}
     \label{log_obs}
\begin{flushleft}
\begin{tabular}{llllrrrrr}
\hline
\hline
\multicolumn{1}{c}{Source} & \multicolumn{1}{l}{Date} & 
\multicolumn{1}{l}{HJDstart} & \multicolumn{1}{c}{Exp.} & 
 \multicolumn{1}{c}{Res.} & \multicolumn{1}{c}{Rate (r')} &
  \multicolumn{1}{r}{Orb. Phase}  & \multicolumn{1}{r}{Obs. mode} & 
 \multicolumn{1}{r}{Filters}  \\
\multicolumn{1}{c}{ } & \multicolumn{1}{c}{yy-mm-dd} & 
\multicolumn{1}{c}{2400000+} & \multicolumn{1}{c}{s} &
\multicolumn{1}{c}{s} & \multicolumn{1}{c}{ct/s} & 
\multicolumn{1}{r}{ }  & \multicolumn{1}{r}{ } & 
\multicolumn{1}{r}{}   \\ 
\hline
\noalign{\smallskip}
VV Pup &	05-05-11 &	53502.47576389	&	1377 &	0.0975 &	14100 &	0.72-0.95 &	Drift  &	 u'  r'        \\
VV Pup &	05-05-11 &	53502.49487269	&	5179 &	0.0975 &	16350 &	0.99-1.85 &	Drift  &	 u', HeII, r' \\
VV Pup &	05-05-13 &	53504.45811343	&	6525 &	0.0975 &	 8414 &	0.14-1.22 &	Drift  &	 u', HeII, r' \\
VV Pup &	05-05-14 &	53504.53365741	&	4885 &	0.0975 &	11150 &	0.22-1.03 &	Drift  &	 u', HeII, r' \\
\noalign{\smallskip}
\hline
\end{tabular}
\end{flushleft}
\end{table*}

The data were reduced using the ULTRACAM pipeline\footnote{http://deneb.astro.warwick.ac.uk/phsaap/software/ultracam/html/}. Corrections for bias and flat field response were applied. An airmass correction was also performed.
To check for the sky and the instrument stability, differential photometry was performed using a nearby non-variable comparison.
No absolute calibration  was applied since we focus on the oscillation variability study. 

\subsection{Orbital light curves}

\begin{figure}
   \centering
    \includegraphics*[width=8.9cm,angle=-0,trim=80 70 40 20]{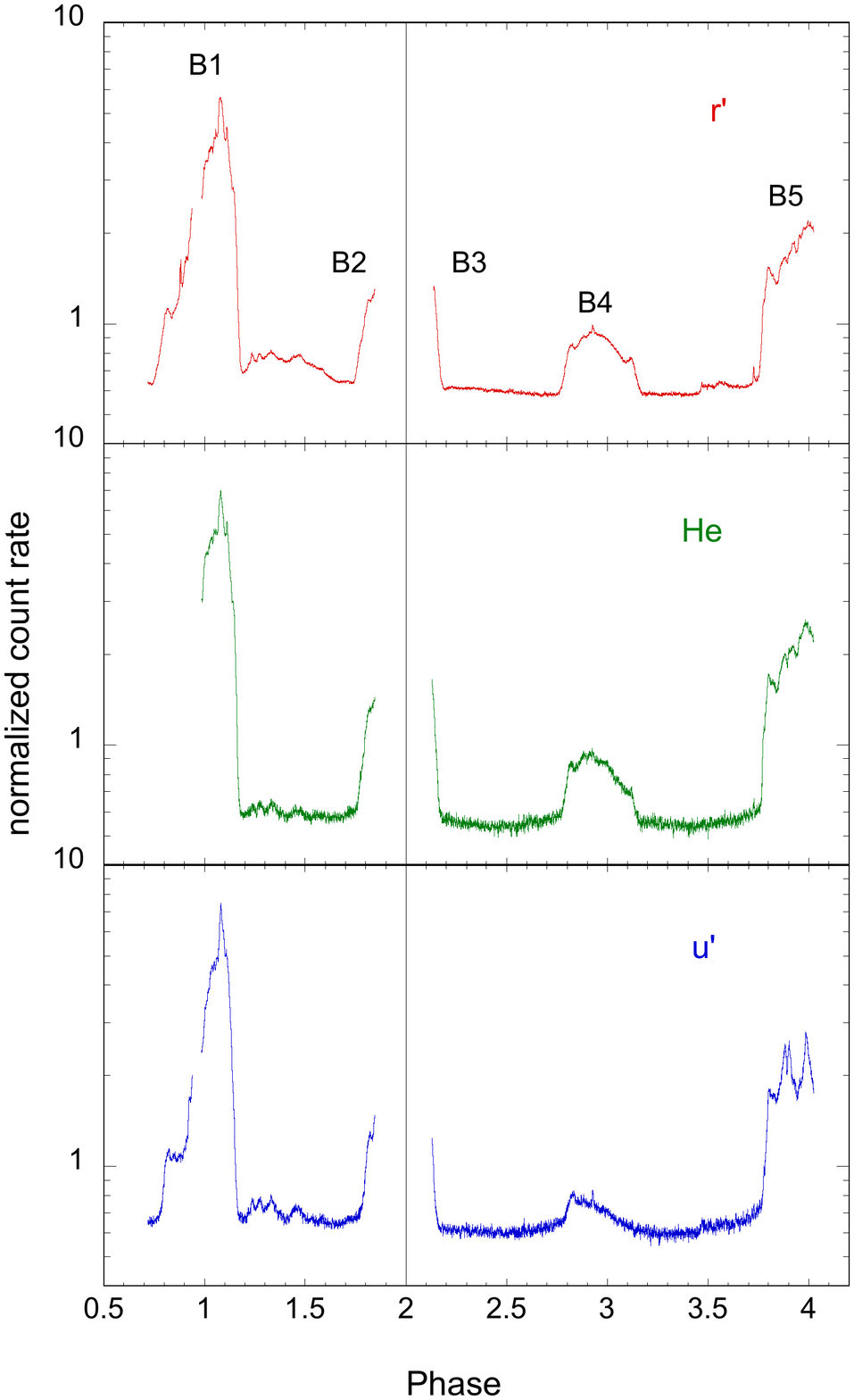}
      \caption{Normalized light curves in the r' (top), He (middle) and u' (bottom) filters versus orbital phase  at a resolution of 9.7 s. 
      The source light curve has been divided by the comparison one. 
      Note that the vertical scale is logarithmic. 
      The observations are spread over four different orbital cycles but split over two different nights (the vertical line indicates the gap of 27 orbital cycles). 
      The different bright phase intervals (B1 to B5) are labelled. 
                }
         \label{FigLC}
   \end{figure}

Figure \ref{FigLC}  shows the light curves for the three filters. The orbital phase is computed using the ephemeris given by 
\citetads{ 1965CoKon..57....1W}, 
where phase 0 refers  to the maximum of the $\sim$1.67 h optical light curve. The separation for the two nights is shown by a vertical line marking a gap of 27 orbital cycles. 
The strong modulation of the light curves clearly indicates that VV Pup was in a high state during the observations. 
The main feature is a bright phase ($\phi$ $\sim 0.75-1.15$) with a smooth rise and rapid drop observed similarly in the three wavelength bands.
A very small bump (at $\phi$ $\sim 1.4$) is visible in the r' and u' bands during the first faint phase which might indicate a weak contribution from a second pole.

Our observations cover five bright phases (labelled B1 to B5) with three only partially covered (B2, B3, and B5). 
Strikingly, the maximum flux in B4 is significantly reduced compared for instance to the B1 maximum by a factor 5.8, 7.4 and 9.1 respectively for the r', He and u' filters. 
Such variability from cycle to cycle was also observed in previous published light curves, though to a lesser extent (see for instance Fig. 1 in
\citeads{ 2000PASP..112...18I}). 

\subsection{QPO characteristics}
 To search for QPOs and study their variability,  we performed Fast Fourier analysis of the light curves in the three filters for consecutive segments of 1024 points  at 0.0975\,s resolution (i.e. 99.8\,s duration), using the {\it powspec} command of the HEASOFT Xronos package\footnote{https://heasarc.gsfc.nasa.gov/xanadu/xronos/xronos.html}. \\
 The significance level of the signal was estimated after normalizing as  defined by 
 \citetads{1983ApJ...266..160L}, 
such that the power distribution is a $\chi^2$ distribution with 2 degrees of freedom (d.o.f.). 
Owing to the additivity of the $\chi^2$  distribution, rebinning the FFTs by averaging W frequency bins and further averaging M individual FFTs will result in a statistical distribution of power according to a  $\chi^2$ distribution with 2WM d.o.f.
For M=1 and W=1, the 99\% confidence detection level corresponds to a power value of 21.7 (see 
\citeads{ 1988tns..conf...27V}). 

\begin{figure}[t]
   \centering
  \includegraphics*[width=8.9cm,height=9.9cm,angle=-0,trim=150 20 70 90]{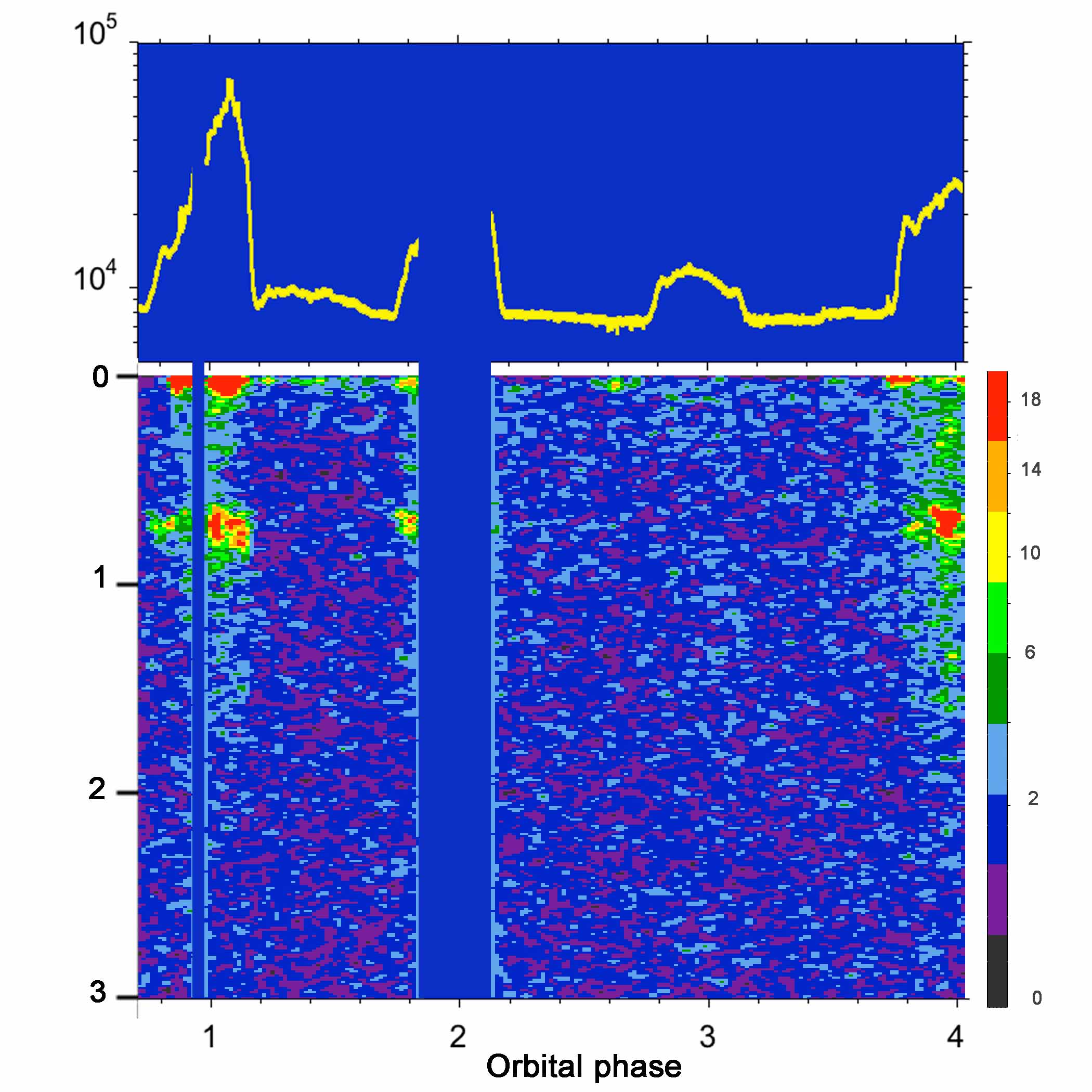}
    \caption{2D Time-Frequency image of the power spectrum for the r' light curve. The X-axis is the orbital phase and the Y-axis is the frequency in the restricted range (0-3)\,Hz. Colour-coded squared amplitude (with \citetads{1983ApJ...266..160L}\,
 normalisation)  is shown according to the scale at right. Individual (99.87s) spectra are shown and the image has been smoothed with a 2-point Gauss filter. The corresponding r' light curve is shown at top.
}
          \label{Fig2Dn}
   \end{figure}

Figure \ref{Fig2Dn}  displays a 2D image of the power spectrum versus orbital phase in the range (0--3) Hz. QPOs around $\sim$ 0.7\,Hz are clearly detected during the bright phases only, with a significant exception during the anomalous B4 bright phase. \\ 
Mean optical QPO amplitudes were computed by averaging FFTs from data in the bright phase ($\phi$ = 0.75--1.15) and faint phase ($\phi$ = 0.15--0.75) intervals and the results are given in Table \ref{optqpo}.  
For the bright phase, amplitudes were evaluated by first subtracting a power-law fitted outside the (0.5--1)\,Hz range and then computing the mean rms values in the (0.5--1)\,Hz interval.
Error bars are evaluated from the effect of adding/subtracting to the power a standard deviation computed in the range (3--5) Hz. 
The mid frequency is given as $f_{med}$, the median value in the range (0.5--1)\,Hz. 
The measured mean amplitudes are of the order of $\sim 1\%$ of the total brightness in the three filters and their frequencies are very similar ($\sim$ 0.74 Hz).
The QPO widths (FWHM) are of the order of 0.11 Hz (r') and up to 0.14 Hz (He, u'). 
Note that the bright phase B4 was excluded as no significant QPOs are detected here.

\begin{table}
\caption[ ]{Mean optical (0.5--1)\,Hz QPO properties.}
     \label{optqpo}
\begin{flushleft}
\begin{tabular}{llcc}
\hline
\hline
\multicolumn{1}{l}{Filter (Rate ct/s)} & 
 \multicolumn{1}{l}{rms $(\%$)} &
  \multicolumn{1}{c}{$f_{med}$} & 
 \multicolumn{1}{c}{FWHM}  \\
  \multicolumn{1}{c}{ }  & 
\multicolumn{1}{l}{  }  & 
\multicolumn{1}{c}{Hz} & 
\multicolumn{1}{c}{Hz} \\
 \hline
\multicolumn{2}{l}{Bright $\phi$=(0.75--1.15)*}  &
\multicolumn{1}{c}{ }  & 
\multicolumn{1}{c}{ }  \\  
\hline
 
 \noalign{\smallskip}
r' (24402)  &   1.04 $\pm 0.06$               &   0.73 $\pm$0.01  &      0.11 $\pm 0.01$  \\
He (3228) &   1.27 $\pm 0.70$                &  0.73 $\pm$0.01   &      0.14 $\pm 0.02$ \\
u' (4542)   &  0.97 $\pm 0.80$                &   0.75  $\pm$0.01  &      0.14 $\pm 0.02$  \\
\hline
\multicolumn{2}{l} {Faint $\phi$=(0.15--0.75)} &  
\multicolumn{1}{c}{ }  & 
\multicolumn{1}{c}{ }  \\
\hline
\noalign{\smallskip}
r' (8239) &   < 0.15 $\pm 0.36$           &-   &  -\\
He (981) &     < 0.57 $\pm 0.96$        &-   &  - \\
u' (1486 ) &    < 0.52   $\pm 0.92$      &-  &  -  \\
\hline
\noalign{\smallskip}
\end{tabular}
\end{flushleft}
(*) excluding B4
\end{table}

For the faint phase, the power in the range (0.5--1)\,Hz was computed after subtracting a 2nd order polynomial continuum fitted outside (0.5--1)\,Hz and error bars were computed as for the bright phase. The upper limits are given in Table  \ref{optqpo} with all amplitudes consistent with zero and best constrained to < 0.15\%  in the r' filter, confirming the absence of QPOs during this phase.

\begin{figure}
   \centering
    \includegraphics*[width=8.9cm,angle=-0,trim=60 70 70 30]{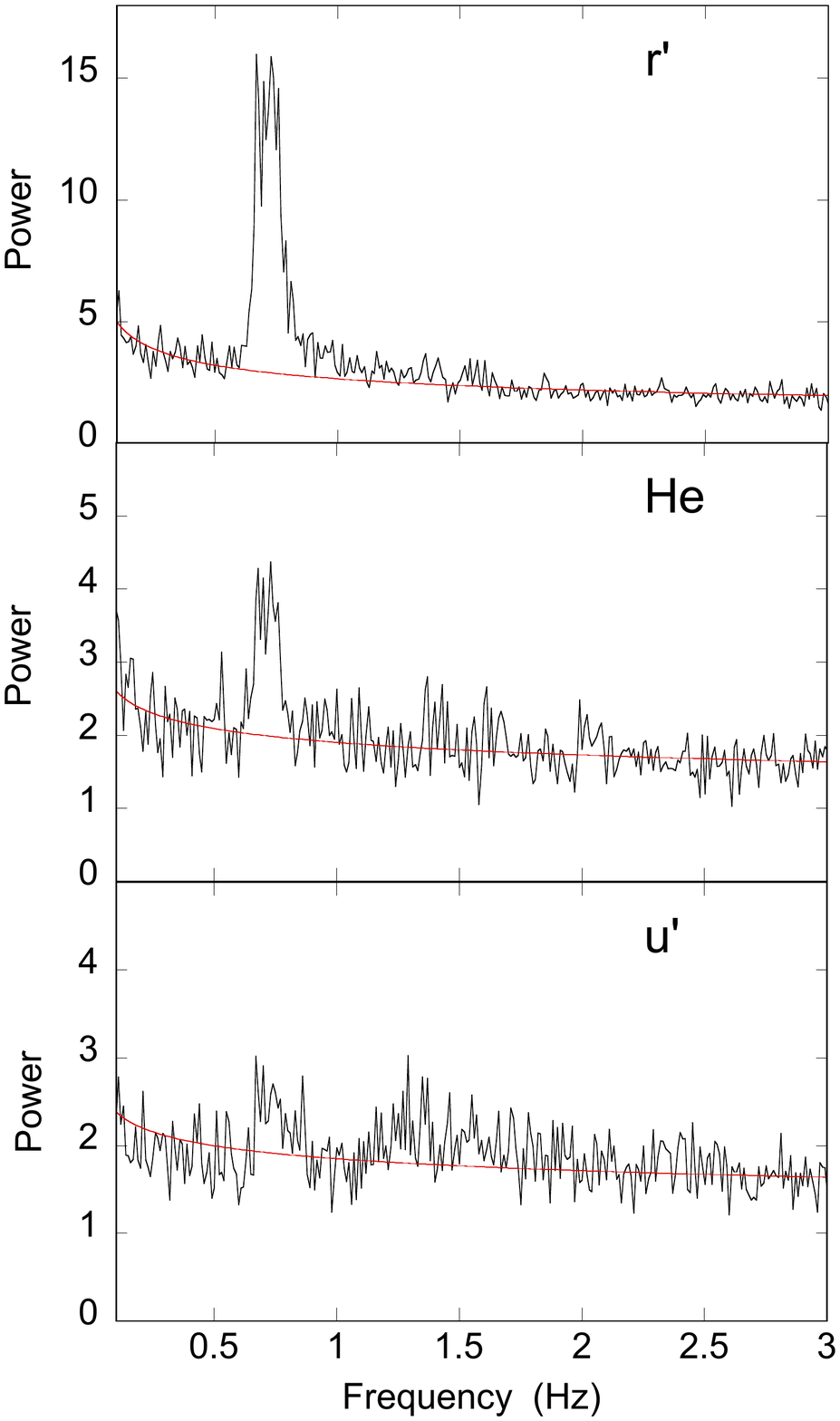}
      \caption{Bright phase QPO frequency profiles. Mean power density spectra are shown in the range (0.1-3)\,Hz for the bright phase interval (0.75-1.15) (excluding B4) in the three wavelength bands  (up: r', middle: He, bottom: u'). The Y scale is the 
      power  computed with Leahy normalization.
      The red line shows the subtracted background fitted as a power-law outside the (0.5-1)\,Hz interval.  }
        \label{FigmeanPSP}
   \end{figure}

\begin{figure}
   \centering
     \includegraphics*[width=8.9cm,height=10.5cm,angle=-0,trim=80 50 50 30]{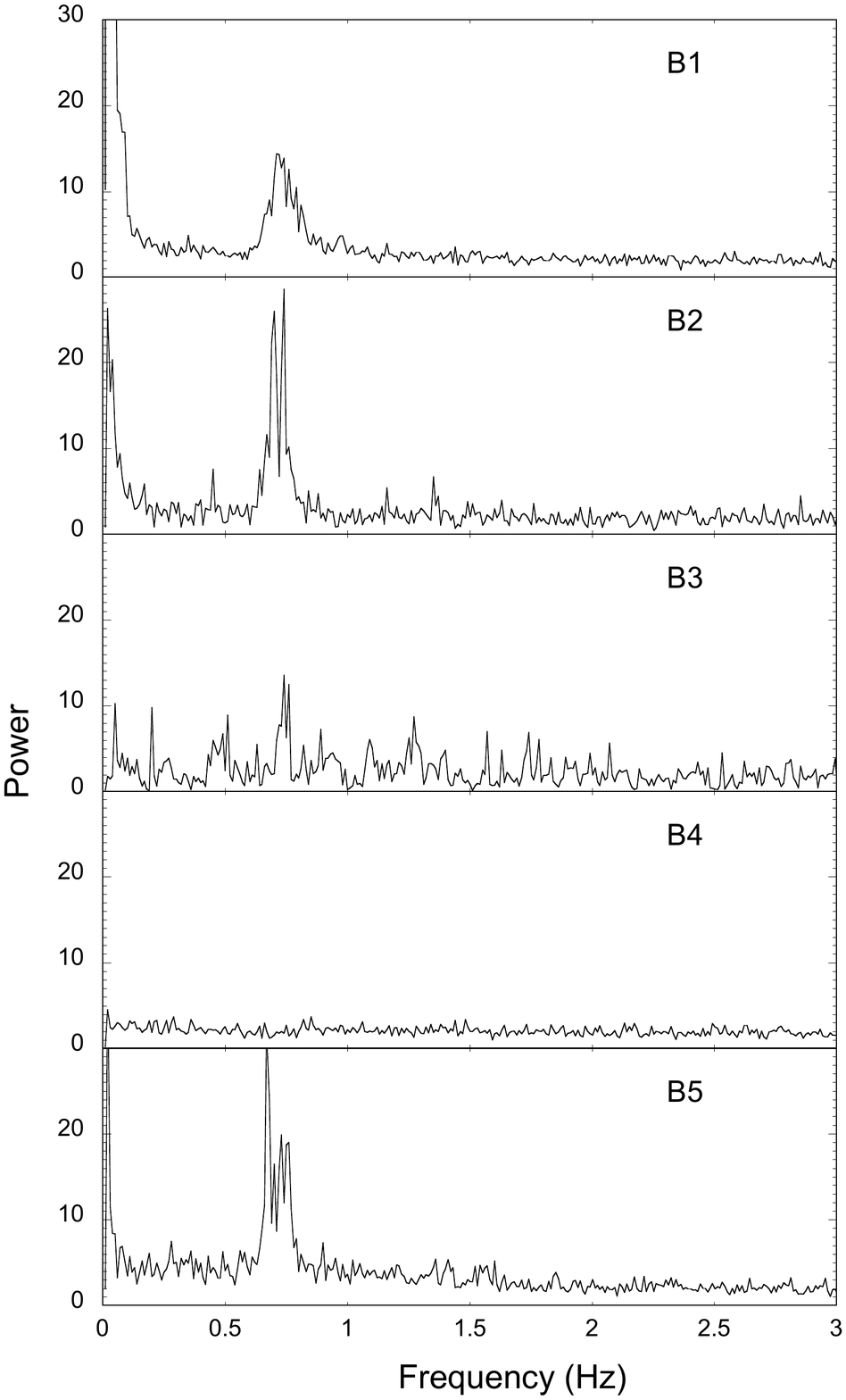}
\caption{Power spectra for the r' light curve in the different bright phases as labelled in Fig. \ref{FigLC}. Note the disappearance of the QPOs in B4.} 
         \label{Fig_Bright5}
   \end{figure}

Figure \ref{FigmeanPSP} shows the QPO frequency profiles averaged during all bright phases (excluding B4), with characteristics given in Table \ref{optqpo}. Besides the excess around $\sim 0.74$\,Hz, there is some hint of additional power around $\sim$ 1.5 Hz in the three filters, most pronounced  in u' though with significant noise variability. \\
We checked the variability in the different orbital cycles by  building a mean FFT for the different bright phases (B1 to B5), shown in Figure \ref{Fig_Bright5} for the r' filter. 
QPOs are well detected in all bright phases even in the short interval B3 but are absent in the bright phase B4. 
The shape of the QPOs is variable, showing clear multiple components in B2 and B5.
The rms amplitudes for the four intervals, B1, B2, B3 and B5, are respectively of 0.86\%, 1.34\%, 0.97\% and 1.39\%. and the upper limit for B4 is <0.40\%. 

   \begin{figure}
   \centering
    \includegraphics*[width=8.9cm,angle=-0,trim=100 100 320 30]{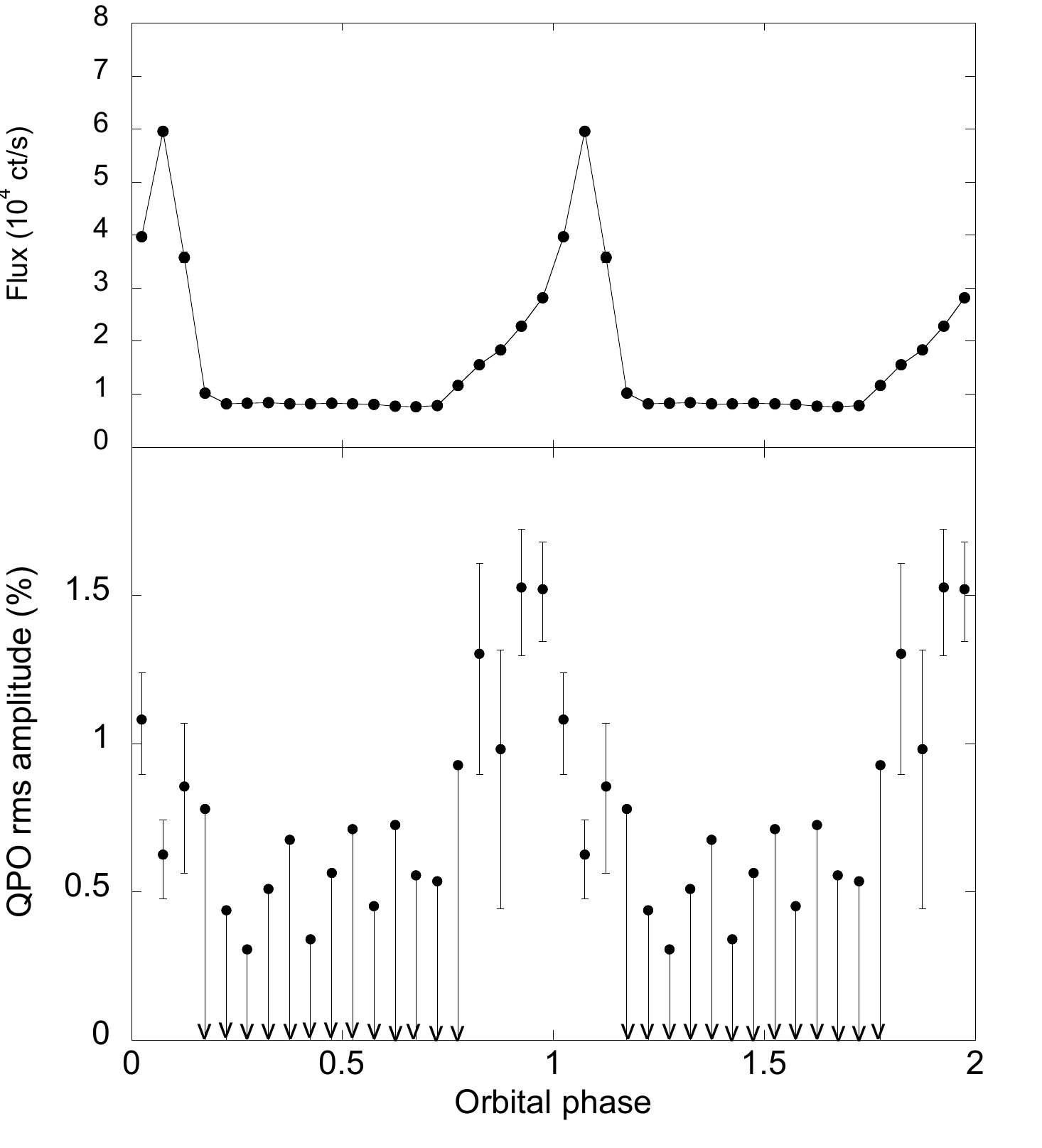}
      \caption{Orbital variation of the QPO rms amplitudes in 0.05 phase intervals for the r' filter (bottom). Only upper limits are derived in the faint phase interval ($(\phi$ = 0.15-0.75). The corresponding r' folded light curve is also shown (top).
                }
         \label{rms_orbphase}
   \end{figure}

\subsection{QPOs orbital and time variability}
\begin{figure*}
     \centering
   \includegraphics*[width=11.5cm,height=8.0cm,angle=-0,trim=70 10 100 60]{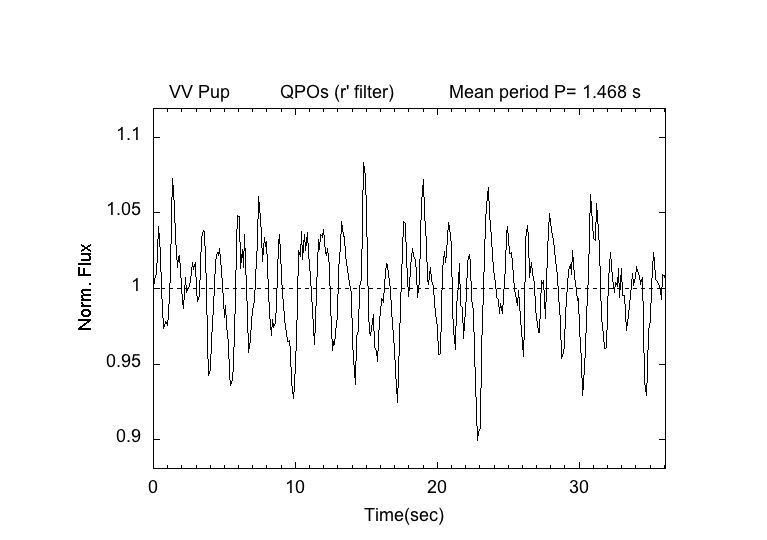}
   \hspace*{0.cm} 
     \includegraphics*[width=6.5cm,height=8.0cm,angle=-0,trim=40 10 40 60]{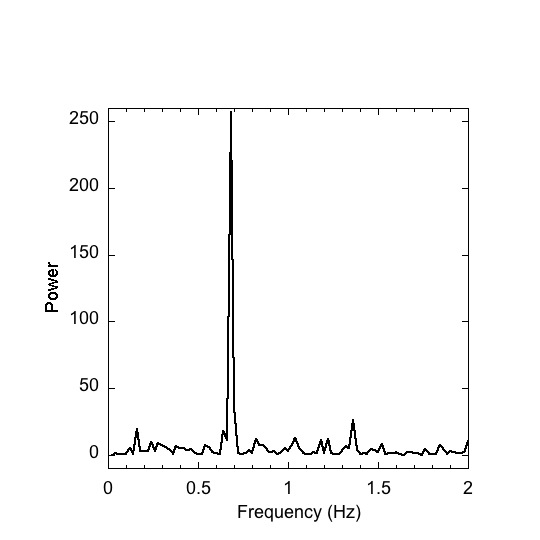}
      \caption{Example of typical VV Pup flux oscillations observed in a 36 s time interval at mid-phase 0.970 (left). The shape and amplitude of the oscillations strongly vary with a mean amplitude of $\sim$ 4\% in a narrow frequency range of $\sim$ 0.68 Hz as shown by the power spectrum (right). 
       }
         \label{Fig_indiv}
   \end{figure*}

To look for QPO orbital modulation, we also computed mean FFTs in 200 phase bins across the orbit.
The measured QPO rms amplitudes are shown in Fig.\ref{rms_orbphase} (lower panel) with the corresponding r' mean counting rate also shown (upper panel).
There is no strict correlation between the relative amplitude and the flux, the QPOs being present at nearly the same relative amplitude ($\sim$ 1-1.5\%) over all the bright phase.  \\
As the  individual quasi-periodic pulsations are seen by eye, we also investigated the QPO characteristics down to short ($\sim$ 50 s) time intervals. As an example, Fig. \ref{Fig_indiv} shows one of the intervals where the QPOs are most prominent with a typical mean amplitude $\sim$ 4\% at a mean period of $\sim$ 1.47\,s.
Inspection of different other intervals clearly reveals that the oscillating signal is strongly variable on a timescale 
of minutes in the same way as already observed in the source V834 Cen (
\citeads{ 2017A&A...600A..53M}). 

\section{XMM-Newton X-ray observations}
VV Pup was observed with the XMM-Newton satellite during a long duration $\sim$ 50 ks exposure on Oct. 20 2007.
The observation in timing mode covers eight consecutive orbital cycles of the source during a high state and has not  yet been fully published. In the context of the analysis of a set of selected polars performed by 
\citetads{ 2015A&A...579A..24B}, 
 an upper limit on fast X-ray QPOs in VV Pup in the range (0.1--5)\,Hz was set at $\sim$28\% during bright phase from this 2007 observation.  
Previous observations have provided only limited information on the source light curve and spectrum based on Einstein, EXOSAT, ROSAT and RXTE satellites
(\citeads{ 1984ApJ...279..785P}, 
\citeads{1985xra..conf...63O}, 
\citeads{ 1996MNRAS.278..285R}, 
\citeads{ 2000PASP..112...18I}). 
An XMM observation of VV Pup is available but is restricted to a typical low state of the source 
\citepads{2005ApJ...620..416P}.  
XMM data were retrieved here to also provide a useful comparison to our optical data.
For our purposes, we restrict the study to the  EPIC-pn data, obtained in timing mode 
\citepads{ 2001A&A...365L..18S}. 
The resulting count rate ($<$10 counts/s) is well below the pile-up limit. Since no high energy ($>$ 10 keV) background flaring is apparent,  the total observation was kept. 
The data were processed using the version V11.0.0  of the Science Analysis Software  (SAS) and FTOOLS software packages\footnote{https://www.cosmos.esa.int/web/xmm-newton/how-to-use-sas}.
We included both single and double events (PATTERN <= 4) in our analysis. 
The barycentric correction was applied for the timing analysis.
The source spectrum was extracted by selecting a 15-column strip around RAWX = 37, corresponding to the brightest signal in the image.
The background spectrum was extracted in columns [4-18], far away from the source.  

\subsection{X-ray light curves}
\begin{figure}
   \centering
   \includegraphics*[width=8.9cm,angle=-0,trim=70 80 40 130]{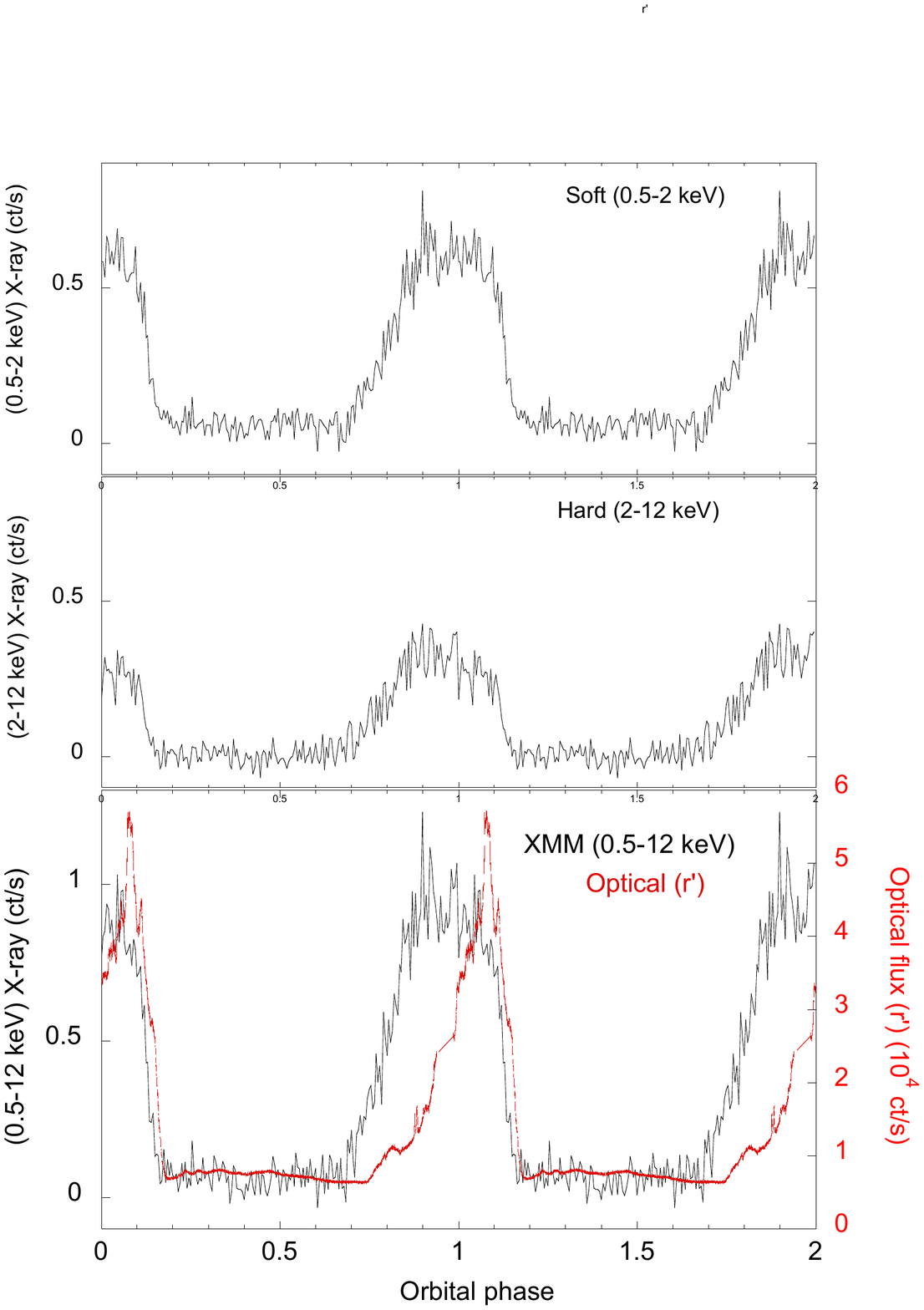}
      \caption{Mean X-ray orbital light curve of VV Pup in the two different bands : soft (0.5--2 keV) (top) and hard (2--12 keV) (middle) and in the full range (0.5--12 keV) (bottom).  The X-ray light curve is an average of 8 consecutive cycles. For comparison, the optical (r') light curve, corresponding to B1, is also shown superposed (in red).
                }
         \label{FigLC_XMM}
   \end{figure}

The total XMM observation covers eight consecutive orbital cycles. Light curves were  produced in the ranges 0.5--2 keV (soft), 2--12 keV (hard) and 0.5--12 keV (full), with resulting mean net count rates of 0.24, 0.11 and 0.35 ct/s respectively. 
Figure \ref{FigLC_XMM} shows the mean X-ray orbital light curves in the three bands, folded with the 0.0697\,d orbital period. 
The mid X-ray maximum occurs at phase 0.953 $\pm 0.003$ (median determination), with respect to the optical maximum as predicted by the Walker's ephemeris
\citepads{ 1965CoKon..57....1W}, 
indicating a very stable orbital/spin period as already noted by 
\citetads{ 2006AJ....131.2216H}.\,
 The  shape of the X-ray bright phase is slightly asymmetrical with a progressive rise and a steeper slope during fall. 
For comparison, the optical (r') light curve corresponding to the B1 cycle (Fig. \ref{FigLC}) is also shown superposed. 
The optical asymmetry is more pronounced and the mid-maximum is at phase 1.047  $\pm 0.005$. 
This significant phase shift was commonly observed revealing that the optical bright phase is variable in position and width 
(\citeads{ 1972MNRAS.156..305W}, 
\citeads{ 1997AN....318..111S}, 
\citeads{ 2000PASP..112...18I}). 
Although the period is reliable over several tens of years, fluctuations of  the accretion flow most probably induce small variations in the location of the accreting region.

\subsection{X-ray QPO search}
In BB15, X-ray QPOs were already searched in the frequency range up to 50 Hz in the (0.5--10 keV) energy range. Here we extend the search in the frequency range up to 125 Hz by using a EPIC-pn time resolution of 0.004 s and analyse the data in the three separate energy ranges as defined above.
To derive upper limits on X-ray QPO amplitudes, we applied the same procedure as in BB15, following the prescription of 
\citetads{ 1988tns..conf...27V}. 

 Averaged power spectra, normalised according to 
  \citetads{1983ApJ...266..160L}, 
 are computed adding individual FFTs of 1024 points intervals.
In none of the average spectra, the maximum power  exceeds the detection limit $P_{detect}$ 
where $P_{detect}$ is the power level that has only a 1\% probability to be exceeded by the noise.
The upper limit is then obtained as $P_{UL}$= $P_{max} - P_{exceed}$ where $P_{max}$ is the highest power in the selected frequency range and $P_{exceed}$ is the power level which will be exceeded by only 1\% of the data in the absence of any periodic signal due to noise.
Results are given in Table \ref{xmmqpo} in the three energy ranges. The relative amplitudes were computed in two frequency ranges (0.1--5)\,Hz and (5--125)\,Hz, assuming  a standard frequency of respectively 1 Hz and 20 Hz. 
The upper limits  for the relative QPO amplitudes  are high ($\sim$30-50 \%), in agreement with values published in BB15.  
As QPOs may be transient, individual power spectra in short 104s intervals were also examined but no positive signal was detected. \\ 

\begin{table*}
\caption[ ]{X-ray fast oscillation (0.1-5) and (5-125) Hz detection limits from XMM-EPIC-pn  observations}
     \label{xmmqpo}
\begin{flushleft}
\begin{tabular}{l l r r r r r r r}
\hline
\hline
\multicolumn{1}{c}{Energy range} & 
 \multicolumn{1}{c}{Time resolution} &
 \multicolumn{1}{c}{Rate} &
  \multicolumn{1}{r}{M}  & 
  \multicolumn{1}{r}{$f_{max}$} & 
 \multicolumn{1}{r}{$P_{max}$}  &
\multicolumn{1}{r}{$P_{exceed}$} & 
 \multicolumn{1}{r}{$P_{detect}$}  &
  \multicolumn{1}{r}{Limit} \\
\multicolumn{1}{c}{ } & 
\multicolumn{1}{c}{ } & 
 \multicolumn{1}{c}{ct/s} & 
\multicolumn{1}{r}{ }  & 
\multicolumn{1}{r}{Hz} & 
\multicolumn{1}{r}{}  & 
\multicolumn{1}{r}{$2.6\,\sigma$} & 
\multicolumn{1}{r}{$2.6\,\sigma$}  & 
\multicolumn{1}{r}{\% rms} \\ \\
\hline
\noalign{\smallskip}
0.5--12 keV & 0.1s &    0.76       &       191    &       3.42   &       2.56   &       1.68    &       2.65    &       29.6    \\
0.5-- 2 keV   & 0.1s &    0.50       &       191    &       2.60   &       2.38   &       1.68    &       2.65    &       29.8    \\
2--12 keV  & 0.1s   &    0.26       &       191    &       0.61   &       2.56   &       1.68    &       2.65    &       41.7    \\
\\
0.5--12 keV & 0.004 s &    0.76       &    4582    &       40.53 &       2.08   &       1.93    &        2.12   &        51.3    \\
0.5--2 keV & 0.004 s &    0.50       &    4582    &       27.10 &       2.06  &       1.93    &        2.12   &        58.9  \\
2--12 keV & 0.004 s &    0.26       &    4582    &       12.45 &       2.06   &       1.93    &        2.12   &        78.5    \\

\noalign{\smallskip}
\hline
\end{tabular}
\end{flushleft}
\end{table*}

\subsection{X-ray energy spectrum}
The VV Pup X-ray energy spectrum was generated using the associated redistribution matrix and ancillary response files and the standard SAS threads. Spectral fits were performed using XSPEC v12.7 HEASOFT software package\footnote{https://heasarc.gsfc.nasa.gov/xanadu/xspec}, in the range (0.5--10 keV) where the calibration accuracy and the count rate statistics are the best.
In the following, all errors are given at a 90$\%$ confidence level for one interesting parameter ($\Delta \chi^2 = 2.71$).

The source spectrum was built from the bright phase only, corresponding to the orbital phase interval (0.75--1.15). We first fitted the data with the most simple X-ray spectrum for polars, consisting of the sum of a blackbody and a bremsstrahlung component, absorbed by cold matter ({\it wabs} model).  
The neutral column density was fixed at $3 \times 10^{19}$ at.cm$^{-2}$, the best fit value derived from EUVE observations 
\citepads{ 1995ApJ...445..921V}. 
An Fe line with Gaussian profile at 6.76 keV was also required in the fit. 
Results are given in Table \ref{xmmspec}. 
The data are best fitted with a 0.14 keV blackbody and a high temperature 39 keV bremsstrahlung, indicating a high white dwarf mass. 
We checked the influence of the  N$_{\rm H}$ value by increasing it by a factor 10, value still lower than the total interstellar one in the direction of the source.
This did not change the blackbody temperature but the bremsstrahlung temperature is slightly lower (33 keV). 

A partial covering  with neutral absorption ({\it pcfabs} model) was also tested, leading to a lower bremsstrahlung temperature (13 keV), a  small covering fraction of  0.27 and a neutral absorption column of 1.0 $ \times 10^{23}$ at.cm$^{-2}$ but without significantly improving the fit (reduced $\chi^2$=1.270 for 133 d.o.f.). 
A reflection of the X-ray continuum onto the surface of the white dwarf can harden the underlying incident spectrum. 
Adding a  convolution of the bremsstrahlung with a {\it reflect} component 
\citepads{ 1995MNRAS.273..837M}, 
 indeed decreases the bremsstrahlung temperature (18 keV) but does not improve the fit significantly ($\chi^2$=1.264 for 134 d.o.f.) and requires a unrealistic large reflection factor of $3.0$.

\begin{figure}
   \centering
    \includegraphics*[width=8.9cm,angle=-0,trim=10 50 90 70]{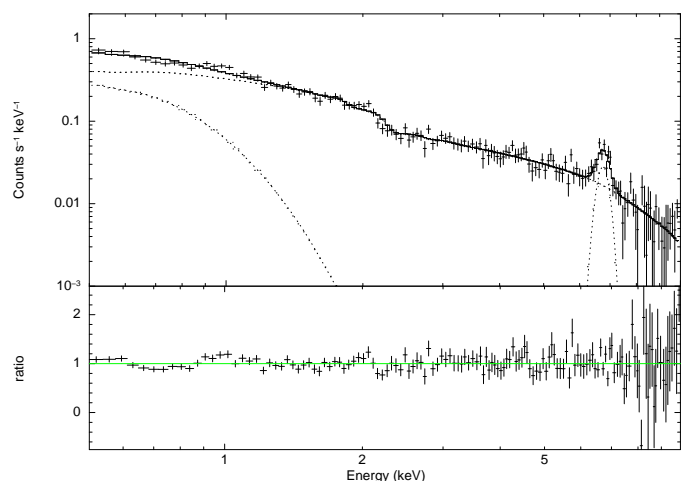}
      \caption{Bright phase X-ray spectrum fitted with the {\it ipolar} model which parameters are given in Table  \ref{xmmspec} (upper frame) and data/model ratio (lower frame). The dotted lines represents the separate component contributions. 
                }
         \label{XMM_spectrum}
   \end{figure}

A more physical approach is described by a stratified post shock region (PSR)  
(\citeads{ 2005A&A...443..291S}, 
\citeyearads{ 2016A&A...591A..35S}). 
The model {\it ipolar} is initially developed for intermediate polars, by specifying the capture point of the material by the magnetic field lines in terms of the white dwarf radius. 
Here we chose to fix the capture radius at 30\,R$_{\rm WD}$
\citepads{1989MNRAS.236...31M},
consistent with a capture inside the Roche lobe of the white dwarf (models are nearly indistinguishable for capture radii greater than 30\,R$_{\rm WD}$).
The results of this fit are given in  Table \ref{xmmspec} and shown in Fig. \ref{XMM_spectrum}.
The best fit gives a high mass compatible with the maximum of 1.4 M$_{\odot}$. 
This is  consistent with the high bremsstrahlung temperature found above.
The blackbody component and the gaussian line are also quite similar to the blackbody-bremsstrahlung fit.
If a neutral partial covering is added to the fit, the white dwarf mass decreases to 0.59 M$_{\odot}$ (with $\chi^2_{red}$=1.255 for 133 d.o.f.), but this value is not well constrained by the fit with a range (0.35 --1.38) M$_{\odot}$ for a 1-$\sigma$ uncertainty.

\begin{table*}
\caption[ ]{Spectral fits of XMM-EPIC-pn data}
     \label{xmmspec}
 \small    
\begin{flushleft}
\begin{tabular}{l l l  |  l l l}
\hline
\hline
\noalign{\medskip}
\multicolumn{3}{c}{Model bb + brems + gaussian} & 
\multicolumn{3}{c}{Model bb + ipolar + gaussian}  \\
\noalign{\medskip}
\hline
\noalign{\smallskip}
\multicolumn{1}{c}{Model component} & 
 \multicolumn{1}{c}{Parameters} &
 \multicolumn{1}{c}{Value} &
\multicolumn{1}{c}{Model component} & 
 \multicolumn{1}{c}{Parameters} &
 \multicolumn{1}{c}{Value}  \\
 \noalign{\smallskip}
\hline
\noalign{\smallskip}
BB & kT$_{bb}$ (keV) &  $0.142_{ -0.018}^{+ 0.017}$ & BB & kT$_{bb}$ (keV) &  $0.138_{ -0.016}^{+ 0.021}$\\
\noalign{\smallskip}
      & norm ($10^{-6}$) & $3.68 $ $_{-0.45}^{+0.50} $ &    & norm. ($10^{-6}$) & $3.55 $ $_{-0.47}^{+0.69} $\\
\noalign{\smallskip}
BREMS           & kT$_{br}$ (keV) &  $39.3_{ -13.5}^{+ 33.2}$ & IPOL  & M$_{WD}$ (M$_{\odot})$ &  $1.40_{ -0.09}^{pegged}$\\
\noalign{\smallskip}
                  & norm ($10^{-4}$) & 5.65  $_{-0.39}^{+0.77} $ &       & norm ($10^{-30}$) & 2.66  $_{-0.072}^{+9.14} $ \\ 
\noalign{\smallskip}
GAUSS & E( keV) & 6.760 $_{ -0.046}^{+ 0.047}$ &  GAUSS & E( keV) & 6.760 $_{ -0.046}^{+ 0.047}$\\
\noalign{\smallskip}
              &  $\sigma$ (keV) & 0.179 $ _{ -0.047}^{+ 0.053}$ &       &  $\sigma$ (keV) & 0.179 $ _{ -0.045}^{+ 0.052}$\\
\noalign{\smallskip}                             
                            & norm ($10^{-5}$) &1.98 $_{-0.42}^{+0.44} $ &   & norm ($10^{-5}$) &1.98  $_{-0.39}^{+0.41} $ \\   
\noalign{\smallskip}   
Flux (0.5-10 keV) & (erg\,s$^{-1}$cm$^{-2}$) & 3.27  $10^{-12}$&  Flux (0.5-10 keV) & (erg\,s$^{-1}$cm$^{-2}$)  & 3.16 $10^{-12}$  \\
\noalign{\smallskip}   
L$_X$ (0.5-10 keV) & (erg\,s$^{-1}$) &       7.4   $10^{30}$  &L$_X$ (0.5-10 keV)& (erg\,s$^{-1}$) &  7.12  $10^{30} $  \\
\noalign{\smallskip}    
 &  $\chi^2_{red}$ (d.o.f)  & 1.261 (135)& & $\chi^2_{red}$ (d.o.f)  & 1.258 (135) \\  
\hline
\end{tabular}
\end{flushleft}
\end{table*}

The results from the different models seem to indicate the presence of a hard component, pointing to a high temperature bremsstrahlung or similarly a high white dwarf mass. However, the restricted XMM energy range is not well suited to confidently evaluate such hard component  which is better constrained by data above 15 keV.
We find that the use of more sophisticated models (neutral partial covering, reflection) gives lower values for  the temperature of the bremsstrahlung or for the WD mass in the {\it ipolar} model, though not significantly improving the fit.
Unfortunately, the VV Pup WD mass is still poorly constrained from X-ray data. A RXTE observation also did not provide useful constraint at high energy since the source was rather faint and only detected up to 5 keV 
\citepads{ 2000PASP..112...18I}.  

Using the precise Gaia distance of 137 $\pm 1$ pc 
(\citeads{ 2018arXiv180409376L}, 
DR2, Gaia catalog), the normalisation coefficients can be used to constrain the accretion geometry.
Assuming typical values of densities and column cross-section (with $n_e=n_i= 10^{16} \, cm^{-3}$ and S=$4\times10^{14}$ cm$^{2}$, see below), the normalisation of the bremsstrahlung component is consistent with a shock height of $h_s= 56$ km.
The normalisation of the minor blackbody component leads to a very small emitting surface (fraction of the white dwarf surface of $2.4 \times 10^{-8}$ for a 0.73 M$_\odot$), 
The blackbody/bremsstrahlung ratio is $\sim$0.05 in the range (0.5--10 keV) and the total unabsorbed (0.5--10 keV) luminosity is $7.4\times10^{30}$ erg\,s$^{-1}$. \\

\section{Discussion}
VV Pup is the archetype of two-pole polars that most of the time shows only one active pole. According to the parameters of the binary system  (i=75$\degr$), the primary southern pole with B = 31 MG is at a colatitude $\beta$ = 145--155$\degr$ and self-eclipsed during a significant part of the orbit 
(\citeads{ 1979ApJ...229..652L}, 
\citeads{ 1989ApJ...342L..35W}, 
\citeads{ 1997AN....318..111S}). 
The white dwarf mass is reasonably well constrained to (0.73$\pm$0.05) M$_{\sun}$ from the secondary orbital velocity measured from detailed infrared spectroscopy during low-state
\citepads{2006AJ....131.2216H},  
and this value will be used in the following.
Although VV Pup has been known for a long time and was the third source classified as a polar 
\citepads{ 1977IAUC.3054....1T}, 
no detailed information on its X-ray light curve and spectrum has yet been published for energies up to 10 keV.
Previous results from earlier satellites were mostly confined to below 5 keV from Einstein satellite 
\citepads{ 1984ApJ...279..785P}, 
EXOSAT
(\citeads{1985xra..conf...63O}, 
\citeads{1985SSRv...40...99M}), 
ROSAT 
\citepads{ 1996MNRAS.278..285R} 
and RXTE  
\citepads{ 2000PASP..112...18I}. 

\subsection{The soft X-ray problem}
Early VV Pup X-ray data from a restricted (0.15--4.0 keV) energy range had pointed to a suspected high soft/hard ratio of  $\sim$15, leading to the concept of a "soft X-ray machine"
\citepads{ 1984ApJ...279..785P}. 
The present XMM data provide the first precise estimation of the hard component. Our best fit indicates a high temperature $\sim$ 39 keV ({\it bremsstrahlung} model) or a massive white dwarf ({\it ipolar} model). We note that the best fitted mass value, close to  the maximum Chandrasekhar mass, is probably an overestimate due to the limited XMM energy range. Hard components (kT > 15 keV) are indeed only weakly constrained in the (0.5--10 keV) range. An additional small soft (0.14 keV) X-ray excess is also required in the fit but contributes negligibly to the total flux. 
The total observed (0.5--10 keV) flux is $3.27\times10^{-12}$ erg.cm$^{-2}.$s$^{-1}$, corresponding to an luminosity of L$_X$ of $7.4\times 10^{30}$ erg.s$^{-1}$ and the total unabsorbed bolometric luminosity will be $2.0\times 10^{31}$ erg.s$^{-1}$.\\
From EUV (0.069--0.18 keV) and soft X-ray (0.1--2 keV) observations, a major soft component was detected 
(\citeads{ 1984ApJ...279..785P}, 
\citeads{ 1996MNRAS.278..285R}, 
\citeads{ 1995ApJ...445..921V}). 
Its temperature is still uncertain in the range of (10--40 eV) with a best fitted value of kT$\sim$ 26 eV from EUVE data and an absorption also poorly constrained in the range (1.9--3.7)$\times 10^{19}$ cm$^{-2}$
\citepads{ 1995ApJ...445..921V}. 
Assuming a blackbody distribution with kT=26 eV and N$_H$=$3\times 10^{19}$ cm$^{-2}$, the EUV flux corresponds to a soft bolometric luminosity of 8.6$\times 10^{31}$ erg.s$^{-1}$  and 3.2$\times 10^{32}$ erg.s$^{-1}$  respectively with and without absorption.\\
The (EUV-soft/XMM-hard) ratio is therefore in the range $\sim$ 4--15,  confirming the major contribution at low energy. We stress however that the EUV-soft emission is still poorly constrained due to uncertainties in the temperature and absorption that have a major impact on the estimated luminosity. 
The soft X-ray excess in polars is thought to arise from the irradiation of the polar cap and/or dense blobs accreted within the white dwarf photosphere
\citepads{1995CAS....28.....W}. 

\subsection{VV Pup : an eclipsing system}
VV Pup exhibits simple X-ray and optical light curves, consisting of a bright phase covering nearly half of the orbital cycle during which the optical and X-ray emission of the lower part of the accretion column above the pole is visible, and a faint phase when this emission region is eclipsed by the body of the white dwarf. The observation of strong linear polarization at the end of the bright phase corroborates this picture
(\citeads{ 1986MNRAS.220..633C}. 
\citeads{ 1990A&A...235..245P}). 

The detailed XMM light curve shows a bright phase of duration ($\Delta \phi= 0.46 \pm 0.02)$ with an asymmetric shape characterized by a slow rise ($\Delta \phi= 0.14 \pm 0.01)$ and an abrupt decline ($\Delta \phi= 0.05 \pm 0.01)$. There is no significant change between the soft (0.5--2 keV) and hard (2--10 keV) light curves. This shape is similarly present in the EUVE range 
\citepads{ 1995ApJ...445..921V}. 
The optical light curve is slightly different with in particular a much more gradual egress ($\Delta \phi \sim 0.25$) but a similar sharp ingress. This is comparable to what has been observed previously, though the optical pole egress is also seen to be shorter in duration in some cases and sometimes more similar to X-rays 
(\citeads{ 1993ApJ...419..793I}, 
\citeyearads{ 2000PASP..112...18I}). 
Noticeable in Fig. \ref{FigLC_XMM}, is also the small shift in phase between the X-ray and optical observations obtained at different dates. Some jitter was already noted by previous authors, indicating small changes in the pole location though, on long timescale, the ephemeris is remarkably stable.

The sharp ingress $\Delta \phi= 0.05 \pm 0.01$ can be used to constrain the size of the emitting region. 
The ingress into self-eclipse corresponds to the interval between the time the column is perpendicular to our field of view and the time when the top of the column disappears behind the line of sight. 
Following 
\citetads{ 1989A&A...217..146L}, 
the ratio h/R$_{WD}$, where h is the shock height delimiting the cyclotron emission region and R$_{WD}$ the white dwarf radius, can be computed from the ingress duration $\Delta \phi$ and source parameters ( i=75$\degr$, $\beta$ = 150$\degr$) as h/R$_{WD}$ $\sim$0.01. For a WD mass of 0.73 M$_{\sun}$, this corresponds to h $\sim$ 76 km. 

\begin{figure}
   \centering
  \includegraphics*[width=8.9cm,angle=-0,trim=20 70 20 50]{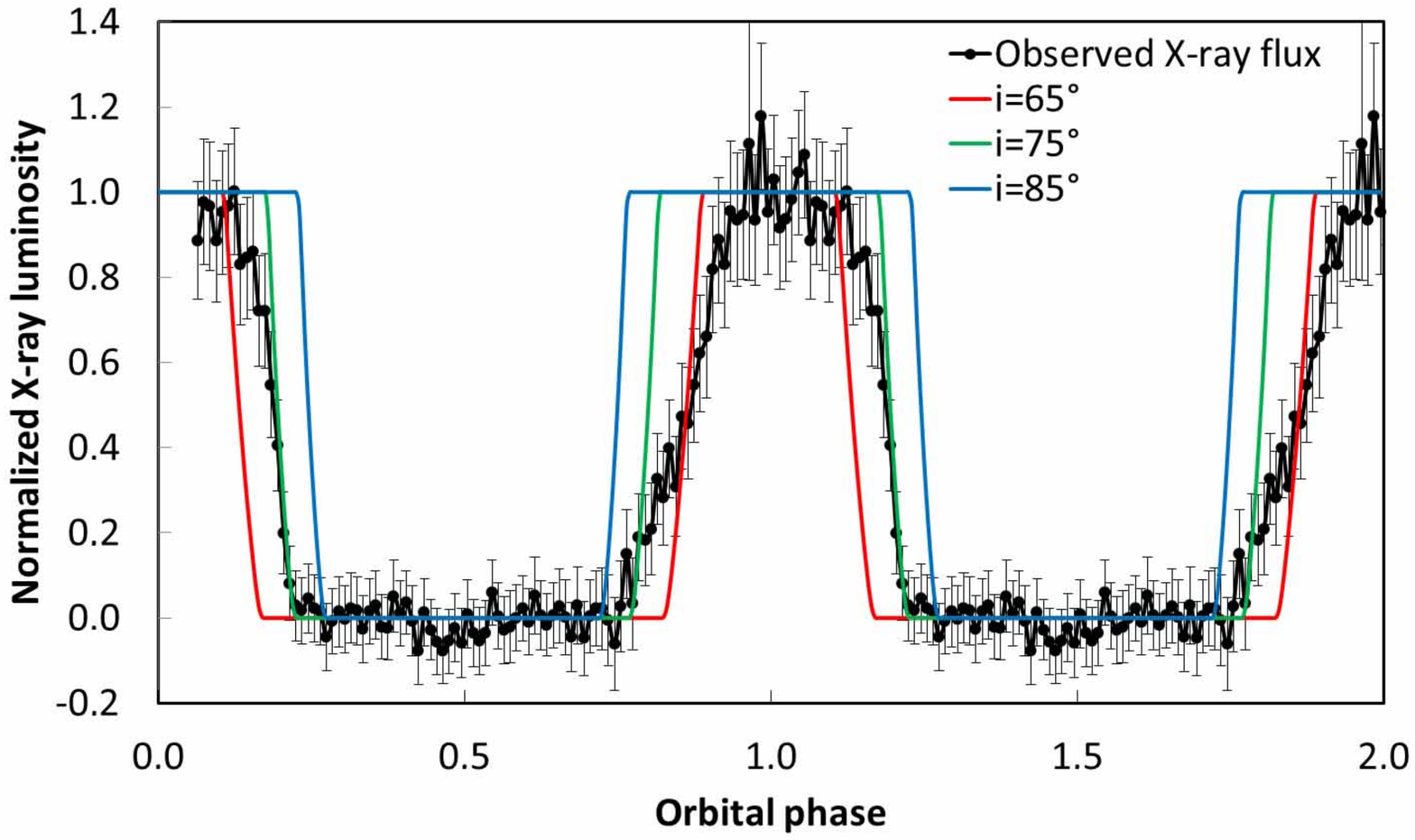}
     \caption{Simulated X-ray light curves according to the source parameters (see text), for different system inclinations (red i=65$\degr$, green 75$\degr$, blue 85$\degr$), compared to  the observed XMM light curve (in black).
                }
         \label{Simu_Xcurve}
   \end{figure}

The hard X-ray flux attributed to the optically thin bremsstrahlung emission of the post-shock region is geometrically modulated by the self eclipse. From a detailed numerical simulation of the post-shock region, X-ray emission can be modelled and typical X-ray light curves can be computed according to the source accretion parameters  (see below \ref{subsec:simulation}).
Figure \ref{Simu_Xcurve} shows the predicted (0.5--10 keV) light curve for a variable inclination and representative source geometrical parameters (colatitude $\beta$ = 150$\degr$, column cross-section S = $4\times10^{14}$ cm$^{2}$). 
The visible part of the accretion column during an orbital phase is evaluated by pure-geometrical effects according to the source parameters (white dwarf and column radius, column height and the $i$ and $\beta$ angles) and the simulated X-ray light curves are derived by integrating the simulated X-ray luminosity through the visible part. Note that here a perfect circular cross-section is assumed for the column.
The general shape is well reproduced, in particular the sharp ingress, for a value $i$ close to 75$\degr$ but not the egress-ingress asymmetry. The more gradual egress may in fact be indicative of a deviation from pure circular shape, pointing to a possible elongated polar spot.
In the same way, the optical flux associated with the optically thick cyclotron emission will also be affected by different projection and limb-darkening effects. We note that the very similar shape of the optical curves in the different filters, with equivalent egress and ingress times, excludes here a pure absorption effect.

%
\subsection{The fast optical oscillations}
VV Pup was the fourth polar found to exhibit optical $\sim 1$\,s QPOs. This feature is not a transient feature and is present during all high states of the source for which observations at  high temporal resolution were performed 
(\citeads{ 1989A&A...217..146L}, 
\citeads{ 1993ApJ...419..793I}, 
\citeyearads{ 2000PASP..112...18I}) 
and is confirmed here with our ULTRACAM observations. 

VV Pup being a pole-eclipse system offers a simple way to directly link the QPOs to the accretion region above the pole. 
Our observations clearly show that optical QPOs are only observed during the restricted bright interval ($\phi$= 0.75--1.15) when the pole region is seen and not in the faint part ($\phi$= 0.15--0.75), confirming previous results
(\citeads{ 1989A&A...217..146L}, 
\citeads{ 1993ApJ...419..793I}). 
The QPOs non-detection during faint phases confirms their link with the cyclotron emission which forms the bulk of the optical bright phase radiation. Note however that the upper limit in faint phase (<0.2--0.5\%) does not preclude the possibility of lower amplitude QPOs that may come from the second pole. 

For the first time, our observations clearly show also that the QPOs may disappear during a full orbital cycle (B4) when the  maximum flux is significantly reduced. As shown in Fig. \ref{Fig_Bright5}, QPOs are however detected in the preceding and following cycles. Such behaviour is not fully explained. In B4, the maximum flux compared to B1 is reduced by a factor $\sim$ 6 and 9, respectively for the red (r') and blue (u') filters. Similar flux variations from cycle to cycle are also apparent in previous observations but with a more reduced factor of $\sim$ 2 (see Fig. 1 in 
\citeads{ 2000PASP..112...18I}). 
One possible explanation might be a temporary lower mass transfer linked to small changes in the capture region, close to the inner
Lagrangian point. In this case, the disappearance of the QPOs will show that they are strongly dependent on the accretion rate. 

The optical QPOs are detected with comparable amplitude $\sim$1\% at the same frequencies of $\sim$0.7 Hz (1.4 s) in the three filters from red to blue and also including the narrow filter HeII (4662 $\AA$). This band was primarily selected to constrain the QPO amplitude in the emission lines (see 
\citeads{2017A&A...600A..53M}).  
However, the underlying continuum is still the dominant fraction of the flux in the filter. The measurement uncertainty prevents an accurate measurement of the contribution of the line flux to the QPOs. However, the fact that the QPO amplitude in this filter is similar to the other bands might indicate that the line flux also contributes to the QPOs.  

The orbital variability (see Fig.\ref{rms_orbphase}) shows that QPOs are clearly detected from the early beginning of the  polar cap egress and show approximatively a constant $\sim$ (1-1.5\%) relative amplitude through all the bright phase till the sharp decrease of the ingress. This is fully in accordance with the visibility of the polar cap and accretion column.

The true nature of the QPOs is not yet settled. They can be the consequence of a naturally broad-band process where different frequencies are excited simultaneously or result from a narrow-band feature varying in frequency with time. Obviously, averaging QPOs over long time intervals will always results in a broad feature, hiding the true QPO nature.
The mean frequency profile, shown in Fig. \ref{FigmeanPSP} with values given in Table \ref{optqpo}, shows that the typical FWHM is $\Delta \nu$ = (0.11--0.14) Hz. For a damped oscillator, this would correspond  to a mean quality factor Q= $\nu$/$\Delta \nu$ = 5--7.
However, for the typical 40 s interval as shown in  Fig. \ref{Fig_indiv}, the quality factor reaches  up to Q=31. In this case, under the assumption of a sine wave shot with an exponentially decaying amplitude with a time constant, the coherence time $\tau$ defined as 
$\tau$ =2Q/$\nu$ is $\sim$ 90 s. 
Such values are higher than what already reported for VV Pup by 
\citetads{ 1993ApJ...419..793I} 
(note however their different definition) and similar to what is also observed in V834 Cen 
\citepads{2017A&A...600A..53M}.  
The amplitudes of these high-coherency intervals are also much higher, reaching more than 5\%. It is therefore more likely that the QPOs are some kind of superposition of these narrow-band features.

\subsection{Hydrodynamical simulations}
  \label{subsec:simulation}
The standard interpretation of the fast QPOs is the hydro-radiative instability of the post-shock accretion region. Their frequencies are indeed characteristic of the timescale of shock oscillations driven by the cooling instability (see the review by
\citeads{2000SSRv...93..611W}).  
Using the multidimensional hydrodynamic code RAMSES 
\citepads{2002A&A...385..337T},  
the radiation hydrodynamics equations can be solved to simulate the radiative accretion dynamics in the accretion column. 
The radiative losses in the accretion column are modeled by a cooling function including both the optically thin bremsstrahlung and the optically thick cyclotron approximated by a power-law
(see 
\citetads{2015A&A...579A..25B} 
and
\citetads{2018MNRAS.473.3158V}  
for details). 

   \begin{figure}
   \centering
\includegraphics*[width=8.9cm,angle=-0,trim=20 0 0 0]{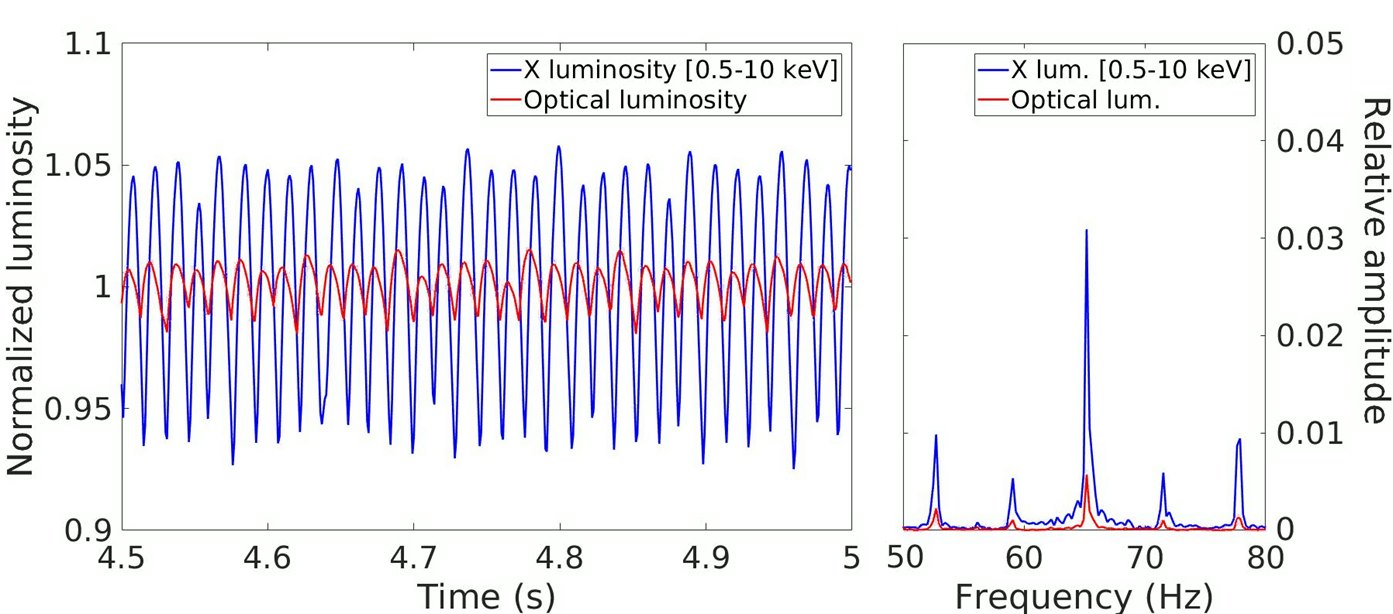}
      \caption{Luminosity oscillations in X-rays (in blue) and optical (in red) predicted by 1D-numerical RAMSES simulations for the VV Pup parameters (M$_{WD}$=$0.73$ M$_{\odot}$, B=31\,MG,  S= $4\times10^{14}$ cm$^{2}$  section and $\dot{\rm M}$ = $2.5\times10^{15}$ g.s$^{-1}$). The corresponding FFTs are shown at right with identical central frequencies at $\sim$ 65 Hz and X-ray and optical amplitudes of 6.8\% and 1.6\% respectively.
       }
         \label{Simu_QPO}
   \end{figure}

From such numerical simulations, relevant synthetic light curves and X-ray spectra can be extracted, depending on four basic source parameters (WD mass, magnetic field B, net accretion rate $\dot{\rm M}$ and accretion cross-section S).
The code was run for a set of parameters relevant to VV Pup with M$_{WD}$=$0.73$ M$_{\odot}$,  B = 31 MG and a total accretion rate of $\dot{\rm M}$ = $0.25\times10^{16}$ g.s$^{-1}$, derived assuming that the total (EUV+soft+hard X-ray) luminosity of L$\rm _x$= $3.4 \times 10^{32}$ erg.s$^{-1}$ corresponds to the gravitational energy of the accreted matter. \\
No direct measure of the column cross-section S is available. 
Only crude estimations can be derived from both spectral and geometrical considerations on the soft X-ray emission, if attributed to the heated hot spot at the basis of the column. From extreme ultraviolet (EUV) and far-ultraviolet (FUV) observations, different estimations are given, ranging respectively from $4\times10^{14}$ cm$^{2}$
\citepads{ 1995ApJ...445..921V} 
to $5\times10^{15}$ cm$^{2}$
\citepads{2002AJ....124.2238H}. 
The size deduced from UV observations is likely to be an overestimate, since UV emission encompasses cooler and larger regions than the limited basis of the accretion column. \\
However, complementary constraints can come from the numerical simulations, if other parameters are fixed, such as the column height. In fact, for a given set of source parameters, varying only the column cross-section results in a monotonic increase of the shock height. For typical VV Pup parameters, a cross-section varying from $2\times10^{14}$  to $10^{15}$ cm$^{2}$ results in an increase of the shock height from 50 to 100 km.
Thereby, assuming a height of \,h $\sim 75$ km as derived from the eclipse indicates a cross-section close to    
S $\sim 4\times10^{14}$ cm$^{2}$ as the best value.

 Figure \ref{Simu_QPO} shows the predicted flux oscillations in X-rays (bremsstrahlung) and optical (cyclotron) for a $4\times10^{14}$ cm$^{2}$ section and a total accretion rate of $2.5\times 10^{15}$ g.s$^{-1}$. Oscillations are present in X-rays and optical with the same dominant frequency of 65.2 Hz and with amplitudes of 6.8\% et 1.6\% respectively. When compared to observations, our upper limits on X-ray oscillations of $\sim$30\% in the range (0.1--5 Hz) and $\sim$50\% in the range (5--125 Hz)  are still only marginally constraining. Also in the optical, the present resolution does not allow to search for predicted frequencies over 5 Hz. However, the optical frequencies detected at (0.6--0.8 Hz) are clearly not predicted by the numerical simulations as exemplified by the comparison of Fig. \ref{Fig_indiv} and Fig. \ref{Simu_QPO}. Therefore, the observations obviously do not confirm the results of the numerical simulations. The same discrepancy was already noted for the source V834 Cen 
\citepads{2017A&A...600A..53M}.  

An extensive study using results from simulations covering a wide range of polar parameters, demonstrates that lower $\sim$ 1 Hz QPO frequencies require a combination of parameters inconsistent with the observed ones
\citepads{2018MNRAS.473.3158V}.  
Namely, for a given white dwarf mass, magnetic field and accretion rate, an increase of the cross-section for instance will have the effect of increasing the cyclotron radiative losses and lowering the frequencies of the oscillations. But in the same time, the amplitudes are also severely reduced to below the detected level, due to the strong damping effect of the cyclotron emission, so that no oscillation will be observed.\\
This is also in accordance with similar 1D simulations that solve the system dynamical evolution using the magnetohydrodynamics code PLUTO 
\citepads{2018MNRAS.474.1629B}.  
In this study, numerical results computed for accretion parameters adapted to the polar V834 Cen, show that  to stay in the observed optical frequency range requires an increase in the column cross-section (in their case a pole radius r$_p$ > 200 km) but with, as a corollary, a negligible oscillation amplitude $\ll 0.1\%$. In this respect, we point out that their conclusions are somewhat misleading since, contrary to what states in their abstract, the characteristics of the post-shock region are not consistent with the observed properties in these simulations. 

We note that some discrepancies exist in their predicted frequencies compared to our recent simulations 
\citepads{2018MNRAS.473.3158V}.  
These differences are due to different assumptions. 
First, for the cyclotron/bremsstrahlung cooling ratio (noted $\epsilon_{s}$), we use the more recent and more rigorous definition given by 
\citetads{1999PhDT........13S},  
more adapted when we take into account the white dwarf gravitation.
Bera et al. (2018) used a cooling function dependent on the column height 
\citepads{1982ApJ...258..289L},   
  that is not appropriate for these numerical simulations.
Also, as the time-dynamical QPOs are not self-similar, the oscillation characteristics depend on the initial conditions. Consequently, taking the steady-state solutions as initial conditions as done in
\citetads{2018MNRAS.474.1629B}  
 introduces a bias. In our simulations, the initial conditions are set by an homogeneous accretion flow striking the white dwarf surface and creating an accretion shock. A non self-similar secondary shock developing near the white dwarf photosphere then starts the oscillations (see 
\citeads{2018MNRAS.473.3158V}).  
The results are therefore non strictly comparable but we consider that our approach is more rigorous.
However, some caution should be taken in the interpretation of the present simulations as they do to incorporate yet detailed boundary conditions such as the possible influence of the soft X-ray emitting polar cap on the incoming accretion flow.

\section{Conclusions}
VV Pup is an interesting polar as it is a pole-eclipse system also showing optical QPOs, allowing more detailed diagnostics : \\ 
-- Optical QPOs are clearly detected here with approximatively the same flux fraction across the visible spectrum, during observations obtained over two different days. However, the observations show for the first time a remarkable interval when QPOs stopped for a full orbital period, during a significant optical flux variation. \\
-- The QPOs are always observed only during the bright phase, indicating that the QPO region is eclipsed and therefore likely linked to the bottom of the accretion column. \\
-- Though the overall mean QPOs appear broadly spread over a (0.6--0.9) Hz interval, there are repetitive time intervals when the QPOs are nearly coherent for durations up to a few minutes, suggesting that the QPO frequency distribution most likely results from a superposition of discrete frequencies. \\
-- Our analysis provides a first estimate of the X-ray spectrum up to 10 keV that indicates a somewhat harder spectrum than expected from the WD mass though this result requires further observations at higher energies. No X-ray QPOs are detected but with an upper limit still too high to put useful constraints on the models.

The observations are still generally in line with what expected from the post-shock region above the WD magnetic pole, the main emitting region. However, important discrepancies remain to be fully in accordance with the model. For the consistent parameters, detailed 1D-simulations failed to reproduce jointly the observed frequencies and amplitudes. As there are no clear alternatives to the shock model, this may point to  important limitations in the treatment of  the boundary conditions above and below the shock region and/or more complex 2D/3D effects
(\citeads{2015A&A...579A..25B}, 
\citeads{2018MNRAS.473.3158V}). 
As the simulations commonly show the presence of oscillations for a rather wide range of parameters, another intriguing point is also the small fraction of polars showing QPOs. This may possibly indicate the need of a fine tuning between the accretion parameters ($\dot{\rm M}$, S). 
The sudden disappearance of the QPOs, observed here from one orbital cycle to the other, may be an example of conditions when such a fine tuning is suppressed due to possible changes in the geometry of the capture region with associated variation in the accretion rate.

Further studies, in the form of 2D/3D numerical simulations and possibly laser astrophysics experiments, are clearly needed before the oscillations observed in polars can be used to provide powerful diagnostics of the WD accretion columns. 
Indeed we have demonstrated that the new powerful lasers devices, as LMJ (Laser MegaJoule, France) and NIF (National Ignition Facility, USA), allow us to reproduce a scaled model of accretion shock 
\citepads{2011ApJ...730...96F}.  
The capability to concentrate around 500 kJ of laser energy in millimetric volume of matter during a few nanoseconds leads to the production of a small-scale accretion column in a similar regime 
\citepads{2018hpl...6..E35}.  
The dynamics, the density and temperature profiles typical of the post-shock region can be achieved and measured with the various diagnostics developed on the facilities (see review in 
\citetads{2006RvMP...78..755R}).  
Preliminary promising results have already been obtained with lasers of intermediate energy 
\citepads{2016NatCo...711899C}  
and current designs are under study for experiments with megajoule lasers (Van Box Som et al. 2019, in preparation). 

\begin{acknowledgements}
Part of this work was supported by the French Programme National de Physique Stellaire (PNPS) of CNRS/INSU co-funded by CEA and CNES. DAHB's research is supported by the National Research Foundation of South Africa. 
\end{acknowledgements}


%
\bibliographystyle{aa}         
\bibliography{VVPup_Ultracam}    

\begin{thebibliography}{49}
\expandafter\ifx\csname natexlab\endcsname\relax\def\natexlab#1{#1}\fi

\bibitem[{{Bera} \& {Bhattacharya}(2018)}]{2018MNRAS.474.1629B}
{Bera}, P. \& {Bhattacharya}, D. 2018, \mnras, 474, 1629

\bibitem[{{Bonnet-Bidaud} {et~al.}(2015){Bonnet-Bidaud}, {Mouchet},
  {Busschaert}, {Falize}, \& {Michaut}}]{2015A&A...579A..24B}
{Bonnet-Bidaud}, J.~M., {Mouchet}, M., {Busschaert}, C., {Falize}, E., \&
  {Michaut}, C. 2015, \aap, 579, A24 [BB15]

\bibitem[{{Busschaert} {et~al.}(2015){Busschaert}, {Falize}, {Michaut},
  {Bonnet-Bidaud}, \& {Mouchet}}]{2015A&A...579A..25B}
{Busschaert}, C., {Falize}, {\'E}., {Michaut}, C., {Bonnet-Bidaud}, J.-M., \&
  {Mouchet}, M. 2015, \aap, 579, A25

\bibitem[{{Chevalier} \& {Imamura}(1982)}]{1982ApJ...261..543C}
{Chevalier}, R.~A. \& {Imamura}, J.~N. 1982, \apj, 261, 543

\bibitem[{{Cropper}(1990)}]{1990SSRv...54..195C}
{Cropper}, M. 1990, \ssr, 54, 195

\bibitem[{{Cropper} \& {Warner}(1986)}]{1986MNRAS.220..633C}
{Cropper}, M. \& {Warner}, B. 1986, \mnras, 220, 633

\bibitem[{{Cross} {et~al.}(2016){Cross}, {Gregori}, {Foster}, {Graham},
  {Bonnet-Bidaud}, {Busschaert}, {Charpentier}, {Danson}, {Doyle}, {Drake},
  {Fyrth}, {Gumbrell}, {Koenig}, {Krauland}, {Kuranz}, {Loupias}, {Michaut},
  {Mouchet}, {Patankar}, {Skidmore}, {Spindloe}, {Tubman}, {Woolsey},
  {Yurchak}, \& {Falize}}]{2016NatCo...711899C}
{Cross}, J.~E., {Gregori}, G., {Foster}, J.~M., {et~al.} 2016, Nature
  Communications, 7, 11899

\bibitem[{{Dhillon} {et~al.}(2007){Dhillon}, {Marsh}, {Stevenson}, {Atkinson},
  {Kerry}, {Peacocke}, {Vick}, {Beard}, {Ives}, {Lunney}, {McLay}, {Tierney},
  {Kelly}, {Littlefair}, {Nicholson}, {Pashley}, {Harlaftis}, \&
  {O'Brien}}]{2007MNRAS.378..825D}
{Dhillon}, V.~S., {Marsh}, T.~R., {Stevenson}, M.~J., {et~al.} 2007, \mnras,
  378, 825

\bibitem[{{Falize} {et~al.}(2011){Falize}, {Michaut}, \&
  {Bouquet}}]{2011ApJ...730...96F}
{Falize}, {\'E}., {Michaut}, C., \& {Bouquet}, S. 2011, \apj, 730, 96

\bibitem[{{Ferrario} {et~al.}(2015){Ferrario}, {de Martino}, \&
  {G{\"a}nsicke}}]{2015SSRv..191..111F}
{Ferrario}, L., {de Martino}, D., \& {G{\"a}nsicke}, B.~T. 2015, \ssr, 191, 111

\bibitem[{{Hoard} {et~al.}(2002){Hoard}, {Szkody}, {Ishioka}, {Ferrario},
  {G{\"a}nsicke}, {Schmidt}, {Kato}, \& {Uemura}}]{2002AJ....124.2238H}
{Hoard}, D.~W., {Szkody}, P., {Ishioka}, R., {et~al.} 2002, \aj, 124, 2238

\bibitem[{{Howell} {et~al.}(2006){Howell}, {Harrison}, {Campbell}, {Cordova},
  \& {Szkody}}]{2006AJ....131.2216H}
{Howell}, S.~B., {Harrison}, T.~E., {Campbell}, R.~K., {Cordova}, F.~A., \&
  {Szkody}, P. 2006, \aj, 131, 2216

\bibitem[{{Imamura} {et~al.}(1993){Imamura}, {Middleditch}, {Scargle},
  {Steiman-Cameron}, {Whitlock}, {Wolff}, \& {Wood}}]{1993ApJ...419..793I}
{Imamura}, J.~N., {Middleditch}, J., {Scargle}, J.~D., {et~al.} 1993, \apj,
  419, 793

\bibitem[{{Imamura} {et~al.}(2000){Imamura}, {Steiman-Cameron}, \&
  {Wolff}}]{2000PASP..112...18I}
{Imamura}, J.~N., {Steiman-Cameron}, T.~Y., \& {Wolff}, M.~T. 2000, \pasp, 112,
  18

\bibitem[{{Langer} {et~al.}(1982){Langer}, {Chanmugam}, \&
  {Shaviv}}]{1982ApJ...258..289L}
{Langer}, S.~H., {Chanmugam}, C., \& {Shaviv}, G. 1982, \apj, 258, 289

\bibitem[{{Langer} {et~al.}(1981){Langer}, {Chanmugam}, \&
  {Shaviv}}]{1981ApJ...245L..23L}
{Langer}, S.~H., {Chanmugam}, G., \& {Shaviv}, G. 1981, \apjl, 245, L23

\bibitem[{{Larsson}(1989)}]{1989A&A...217..146L}
{Larsson}, S. 1989, \aap, 217, 146

\bibitem[{{Leahy} {et~al.}(1983){Leahy}, {Darbro}, {Elsner}, {Weisskopf},
  {Kahn}, {Sutherland}, \& {Grindlay}}]{1983ApJ...266..160L}
{Leahy}, D.~A., {Darbro}, W., {Elsner}, R.~F., {et~al.} 1983, \apj, 266, 160

\bibitem[{{Liebert} \& {Stockman}(1979)}]{1979ApJ...229..652L}
{Liebert}, J. \& {Stockman}, H.~S. 1979, \apj, 229, 652

\bibitem[{{Luri} {et~al.}(2018){Luri}, {Brown}, {Sarro}, {Arenou},
  {Bailer-Jones}, {Castro-Ginard}, {de Bruijne}, {Prusti}, {Babusiaux}, \&
  {Delgado}}]{2018arXiv180409376L}
{Luri}, X., {Brown}, A.~G.~A., {Sarro}, L.~M., {et~al.} 2018, ArXiv e-prints
  1804.09376

\bibitem[{{Magdziarz} \& {Zdziarski}(1995)}]{1995MNRAS.273..837M}
{Magdziarz}, P. \& {Zdziarski}, A.~A. 1995, \mnras, 273, 837

\bibitem[{{Mason} {et~al.}(2007){Mason}, {Wickramasinghe}, {Howell}, \&
  {Szkody}}]{2007A&A...467..277M}
{Mason}, E., {Wickramasinghe}, D., {Howell}, S.~B., \& {Szkody}, P. 2007, \aap,
  467, 277

\bibitem[{{Mason}(1985)}]{1985SSRv...40...99M}
{Mason}, K.~O. 1985, \ssr, 40, 99

\bibitem[{{Meggitt} \& {Wickramasinghe}(1989)}]{1989MNRAS.236...31M}
{Meggitt}, S.~M.~A. \& {Wickramasinghe}, D.~T. 1989, \mnras, 236, 31

\bibitem[{{Mouchet} {et~al.}(2017){Mouchet}, {Bonnet-Bidaud}, {Van Box Som},
  {Falize}, {Buckley}, {Breytenbach}, {Ashley}, {Marsh}, \&
  {Dhillon}}]{2017A&A...600A..53M}
{Mouchet}, M., {Bonnet-Bidaud}, J.-M., {Van Box Som}, L., {et~al.} 2017, \aap,
  600, A53

\bibitem[{{Mukai}(2017)}]{2017PASP..129f2001M}
{Mukai}, K. 2017, \pasp, 129, 062001

\bibitem[{{Osborne} {et~al.}(1985){Osborne}, {Mason}, {Bonnet-Bidaud},
  {Beuermann}, \& {Rosen}}]{1985xra..conf...63O}
{Osborne}, J., {Mason}, K.~O., {Bonnet-Bidaud}, J.~M., {Beuermann}, K., \&
  {Rosen}, S. 1985, in X-ray Astronomy '84, ed. M.~{Oda} \& R.~{Giacconi},
  63--66

\bibitem[{{Pandel} \& {C{\'o}rdova}(2005)}]{2005ApJ...620..416P}
{Pandel}, D. \& {C{\'o}rdova}, F.~A. 2005, \apj, 620, 416

\bibitem[{{Patterson} {et~al.}(1984){Patterson}, {Beuermann}, {Lamb},
  {Fabbiano}, {Raymond}, {Swank}, \& {White}}]{1984ApJ...279..785P}
{Patterson}, J., {Beuermann}, K., {Lamb}, D.~W., {et~al.} 1984, \apj, 279, 785

\bibitem[{{Piirola} {et~al.}(1990){Piirola}, {Coyne}, \&
  {Reiz}}]{1990A&A...235..245P}
{Piirola}, V., {Coyne}, G.~V., \& {Reiz}, A. 1990, \aap, 235, 245

\bibitem[{{Ramsay} {et~al.}(1996){Ramsay}, {Cropper}, \&
  {Mason}}]{1996MNRAS.278..285R}
{Ramsay}, G., {Cropper}, M., \& {Mason}, K.~O. 1996, \mnras, 278, 285

\bibitem[{{Remington} {et~al.}(2006){Remington}, {Drake}, \&
  {Ryutov}}]{2006RvMP...78..755R}
{Remington}, B.~A., {Drake}, R.~P., \& {Ryutov}, D.~D. 2006, Reviews of Modern
  Physics, 78, 755

\bibitem[{{Saxton}(1999)}]{1999PhDT........13S}
{Saxton}, C.~J. 1999, PhD thesis, Univ.~Sydney, Australia

\bibitem[{{Schwope} \& {Beuermann}(1997)}]{1997AN....318..111S}
{Schwope}, A.~D. \& {Beuermann}, K. 1997, Astronomische Nachrichten, 318, 111

\bibitem[{{Str{\"u}der} {et~al.}(2001){Str{\"u}der}, {Briel}, {Dennerl},
  {Hartmann}, {Kendziorra}, {Meidinger}, {Pfeffermann}, {Reppin}, {Aschenbach},
  {Bornemann}, {Br{\"a}uninger}, {Burkert}, {Elender}, {Freyberg}, {Haberl},
  {Hartner}, {Heuschmann}, {Hippmann}, {Kastelic}, {Kemmer}, {Kettenring},
  {Kink}, {Krause}, {M{\"u}ller}, {Oppitz}, {Pietsch}, {Popp}, {Predehl},
  {Read}, {Stephan}, {St{\"o}tter}, {Tr{\"u}mper}, {Holl}, {Kemmer}, {Soltau},
  {St{\"o}tter}, {Weber}, {Weichert}, {von Zanthier}, {Carathanassis}, {Lutz},
  {Richter}, {Solc}, {B{\"o}ttcher}, {Kuster}, {Staubert}, {Abbey}, {Holland},
  {Turner}, {Balasini}, {Bignami}, {La Palombara}, {Villa}, {Buttler},
  {Gianini}, {Lain{\'e}}, {Lumb}, \& {Dhez}}]{2001A&A...365L..18S}
{Str{\"u}der}, L., {Briel}, U., {Dennerl}, K., {et~al.} 2001, \aap, 365, L18

\bibitem[{{Suleimanov} {et~al.}(2016){Suleimanov}, {Doroshenko}, {Ducci},
  {Zhukov}, \& {Werner}}]{2016A&A...591A..35S}
{Suleimanov}, V., {Doroshenko}, V., {Ducci}, L., {Zhukov}, G.~V., \& {Werner},
  K. 2016, \aap, 591, A35

\bibitem[{{Suleimanov} {et~al.}(2005){Suleimanov}, {Revnivtsev}, \&
  {Ritter}}]{2005A&A...443..291S}
{Suleimanov}, V., {Revnivtsev}, M., \& {Ritter}, H. 2005, \aap, 443, 291

\bibitem[{{Tapia}(1977)}]{1977IAUC.3054....1T}
{Tapia}, S. 1977, \iaucirc, 3054

\bibitem[{{Teyssier}(2002)}]{2002A&A...385..337T}
{Teyssier}, R. 2002, \aap, 385, 337

\bibitem[{{Van Box Som} {et~al.}(2018{\natexlab{a}}){Van Box Som}, {Falize},
  {Bonnet-Bidaud}, {Mouchet}, {Busschaert}, \& {Ciardi}}]{2018MNRAS.473.3158V}
{Van Box Som}, L., {Falize}, {\'E}., {Bonnet-Bidaud}, J.-M., {et~al.}
  2018{\natexlab{a}}, \mnras, 473, 3158

\bibitem[{{Van Box Som} {et~al.}(2018{\natexlab{b}}){Van Box Som}, {Falize},
  {Koenig}, {Sakawa}, {Albertazzi}, {Barroso}, {Bonnet-Bidaud}, {Busschaert},
  {Ciardi}, {Hara}, {Katsuki}, {Kumar}, {Lefevre}, {Michaut}, {Michel},
  {Miura}, {Morita}, {Mouchet}, {Rigon}, {Sano}, {Shiiba}, {Shimogawara}, \&
  {Tomiya}}]{2018hpl...6..E35}
{Van Box Som}, L., {Falize}, E., {Koenig}, M., {et~al.} 2018{\natexlab{b}},
  High Power Laser Science and Engineering, 6, E35

\bibitem[{{van der Klis}(1988)}]{1988tns..conf...27V}
{van der Klis}, M. 1988, in Timing Neutron Stars, eds. H. Ogelman and E.P.J.
  van den Heuvel. NATO ASI Series C, Vol. 262, p. 27-70. Dordrecht: Kluwer,
  1988., 27--70

\bibitem[{{van Gent}(1931)}]{1931BAN.....6...93V}
{van Gent}, H. 1931, \bain, 6, 93

\bibitem[{{Vennes} {et~al.}(1995){Vennes}, {Szkody}, {Sion}, \&
  {Long}}]{1995ApJ...445..921V}
{Vennes}, S., {Szkody}, P., {Sion}, E.~M., \& {Long}, K.~S. 1995, \apj, 445,
  921

\bibitem[{{Walker}(1965)}]{1965CoKon..57....1W}
{Walker}, M.~F. 1965, Commmunications of the Konkoly Observatory Hungary, 57, 1

\bibitem[{{Warner}(1995)}]{1995CAS....28.....W}
{Warner}, B. 1995, Cambridge Astrophysics Series, 28

\bibitem[{{Warner} \& {Nather}(1972)}]{1972MNRAS.156..305W}
{Warner}, B. \& {Nather}, R.~E. 1972, \mnras, 156, 305

\bibitem[{{Wickramasinghe} {et~al.}(1989){Wickramasinghe}, {Ferrario}, \&
  {Bailey}}]{1989ApJ...342L..35W}
{Wickramasinghe}, D.~T., {Ferrario}, L., \& {Bailey}, J. 1989, \apjl, 342, L35

\bibitem[{{Wu}(2000)}]{2000SSRv...93..611W}
{Wu}, K. 2000, \ssr, 93, 611

\end{thebibliography}

\end{document}